\documentclass[11pt]{article}
\usepackage{fullpage}
\usepackage{graphicx} 
\usepackage[colorlinks=true,allcolors=blue,breaklinks,draft=false]{hyperref}   
\usepackage{algorithm}
\usepackage{subcaption}
\usepackage[noend]{algpseudocode}
\usepackage{multicol}
\usepackage{authblk} 

\newtheorem{invariant}{Invariant}

\newcommand{\inred}[1]{{\color{red}{ #1 }}}
\newcommand{\idit}[1]{{\color{red}{[\textbf{Idit:} #1 ]}}}

\newcommand{\remove}[1]{}
\newcommand{\code}[1]{\texttt{#1}}

\title{KV-Tandem -- a Modular Approach \\ to Building High-Speed LSM Storage Engines}

\author[1]{Edward Bortnikov, Michael Azran, Asa Bornstein, Shmuel Dashevsky, Dennis Huang, Omer Kepten, Michael Pan, Gali Sheffi, Moshe Twitto, Tamar Weiss Orzech\thanks{Email: \{ebortnik,michaela,asib,shmueld,dennish,omerk,michaelp,galis,moshet,tamarw\}@pliops.com. 
This research was performed while the authors were at Pliops.}}
\affil[1]{Pliops} 

\author[2]{Idit Keidar\thanks{Email: idish@ee.technion.ac.il}}
\affil[2]{Technion ECE}

\author[3]{Guy Gueta\thanks{Email: guy.gueta@intel.com}}
\affil[3]{Intel}

\author[4]{Roey Maor\thanks{Email: roeym@meta.com}}
\affil[4]{Meta}

\author[5]{Niv Dayan\thanks{Email: nivdayan@cs.toronto.edu}}
\affil[5]{Toronto University CS}

\date{}

\begin{document}

\maketitle

\begin{abstract}
We present~\emph{KV-Tandem}, a modular architecture for building LSM-based storage engines on top of simple, non-ordered persistent key-value stores (KVSs). KV-Tandem enables advanced functionalities such as range queries and snapshot reads, while maintaining the native KVS performance for random reads and writes. Its modular design offers better performance trade-offs compared to previous KV-separation solutions, which struggle to decompose the monolithic LSM structure. Central to KV-Tandem is~\emph{LSM bypass} -- a novel algorithm that offers a fast path to basic operations while ensuring the correctness of advanced APIs.

We implement KV-Tandem in \emph{XDP-Rocks}, a RocksDB-compatible storage engine that leverages the XDP KVS and incorporates practical design optimizations for real-world deployment. Through extensive microbenchmark and system-level comparisons, we demonstrate that XDP-Rocks achieves 3x to 4x performance improvements over RocksDB across various workloads. XDP-Rocks is already deployed in production, delivering significant operator cost savings consistent with these performance gains.

\end{abstract}

\section{Introduction}
\subsection{Motivation and goal}
Popular \emph{ordered key-value storage engines} such as RocksDB and its siblings~\cite{RocksDB, Terarkdb, BlobDB, Titan, Pebble}   
provide access to values associated with unique keys. This includes not only point put/get queries of individual keys, but also snapshots (allowing users to read multiple values at a common point in time) and iterators. 
Such state-of-the-art systems are typically organized as \emph{log-structured merge (LSM)} trees~\cite{LSM96}, LSMs for short, which facilitate ordered operations such as range queries. However, despite their advantages for ordered operations, LSM-based systems incur significant trade-offs in write, read, and space efficiency.

In a nutshell, LSMs buffer written data  in memory, periodically flushing it to disk as sorted files and merging those files into larger ones through a  background \emph{compaction} process. Yet this multi-file structure  introduces three key inefficiencies: \emph{write amplification (WA)}, as data is rewritten (copied to a different file) multiple times; \emph{read amplification (RA)}, due to searches across multiple files; and  \emph{space amplification (SA)}, as storage is used inefficiently until compaction removes stale entries.
These three performance metrics are at inherent tension with each other~\cite{LuoCarey}, and hinder the performance of existing systems, particularly in write-heavy workloads, leading to high tail latency, limited scalability, or significant space consumption~\cite{DayanI18, DayanWDPBT22}. 

\remove{
The key challenge hampering such systems is the need for updates to maintain the elaborate functionality (e.g., the order) concurrently with ongoing queries.
Techniques like \emph{key-value separation}~\cite{wiscKey} are useful for mitigating this difficulty, but, as of today, suffer from a number of shortcomings that preclude their widespread deployment, as we explain below.
}

In contrast, an \emph{unordered key-value store (KVS)}, (e.g., a SNIA-defined KVSSD~\cite{KVSSD}) provides efficient point put/get access without supporting complex queries like snapshots or range scans. This allows for lightweight solutions based on log-structured storage indexed by an in-memory hashtable, e.g., Bitcask~\cite{Bitcask} and Pliops's 
\emph{Extreme Data Processor (XDP)}~\cite{DayanI18},
which achieve higher throughput and lower WA. However, unordered KVSs lack support for ordered operations (unless all the data can be indexed in memory).

\remove{
. Unlike an ordered storage engine, a KVS offers only point put/get access. Forgoing support for snapshots and range queries allows for lightweight solutions based on log-structured storage indexed by an in-memory hashtable, e.g., Bitcask~\cite{Bitcask}. A state-of-the-art KVS, e.g., the Pliops Extreme Data Processor (XDP)~\cite{DayanI18}, can serve random reads with a single I/O and offers a write throughput close the the underlying storage device's. For comparison, a highly optimized LSM  entails  double-digit WA, thereby exploiting only a small percentage of the write bandwidth~\cite{LuoCarey}. 
On the negative side, unordered KVSs 
do not support range queries (except in DRAM-bounded data stores where all the data can be indexed in memory~\cite{ForestDB}). While common NoSQL workloads use range queries much less frequently than random access~\cite{SrinivasanBCSGI16}, applications do need to use them from time to time, e.g., for sorted set listings~\cite{RedisAPI} or administrative tasks like whole-database backup~\cite{RoF}.
}
 
The aim of this work is to get the best of both worlds, namely, to serve point gets and puts at (almost) maximum KVS speed with minimal SA, and to offer alongside them the full functionality of RocksDB-like LSM storage engines. We target common NoSQL workloads, which use range queries much less frequently than random  access~\cite{SrinivasanBCSGI16}, but do use them from time to time, e.g., for sorted set listings~\cite{RedisAPI} or administrative tasks like whole-database backup~\cite{RoF}.
Such workloads are common among consumer Internet application providers, e.g., social networks~\cite{CaoDVD20}, including some of our  customers.


\subsection{KV-Tandem}
\begin{figure}
    \centering
    \includegraphics[width=0.5\columnwidth]{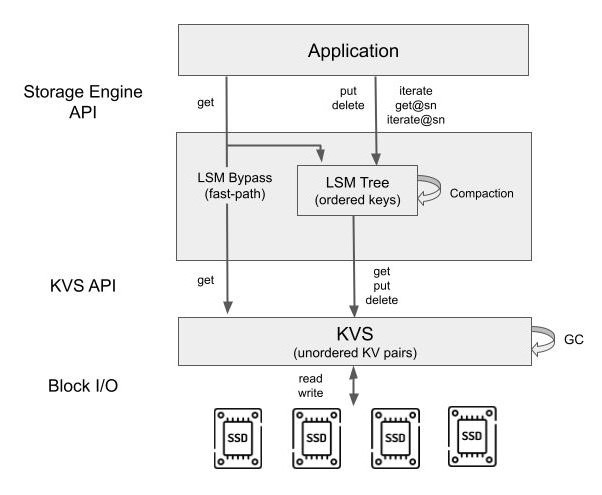}
    \caption{KV-Tandem high-level architecture: providing a storage engine API (RocksDB-compatible) atop a KVS. Values are embedded in the KVS directly, while their keys are embedded indirectly through the LSM tree. The LSM bypass offers a fast-path for point queries, while iterators and snapshots go through the LSM. The KVS GC and the LSM compaction are independent.}
    \label{fig:XDP-Rocks}
\end{figure}

In this work we identify and exploit KVS as a fundamental building block for a feature-rich high-throughput ordered storage engine. Our design, \emph{KV-Tandem}, couples two distinct paradigms. Its unordered value storage is a KVS, whereas its ordered key storage is an LSM. A novel \emph{LSM bypass} mechanism allows point queries to 
directly access the KVS without going through the LSM. The LSM itself 
is also stored as a collection of KV-pairs in the same KVS, simplifying storage management. The KV-Tandem architecture is depicted in Figure~\ref{fig:XDP-Rocks}. This design is generic and agnostic to the KVS's implementation (be it in software or hardware, operating on directly-attached or remote storage). 

Restricting the LSM to store keys (without values) leverages the \emph{key-value separation} approach pioneered in  WiscKey~\cite{wiscKey}, which
 eliminates the bulk of the WA cost by avoiding repeated compaction of large values. Since keys are often much smaller than values (see, e.g., a  survey of Meta workloads~\cite{CaoDVD20}), this approach greatly reduces compaction overhead.

A number of  works, including production storage engines~\cite{Terarkdb, Titan, BlobDB}, have adopted and enhanced key-value separation in different ways. The idea is to  store values in log files, with the LSM holding physical pointers into these log files. However, these systems only achieve competitive LSM read performance when all keys fit in memory~\cite{wiscKey}. Otherwise, queries suffer from additional indirection I/Os due to first searching the LSM for a given key and then chasing pointers to the log files. They also typically induce redundant LSM accesses for value-log \emph{garbage collection (GC)}, hampering write throughput~\cite{wiscKey}. Finally and most problematically, some of these solutions~\cite{BlobDB,Terarkdb} incur high (unbounded) SA under write-intensive workloads, caused by tight coupling between the GC and the LSM's compaction.

We, instead, treat unordered key-value storage as a first-class citizen, and embody it through a state-of-the-art KVS. The latter handles search, update, and GC internally through scalable mechanisms tailored to the simple KVS API. We complement the KVS with an LSM, which in a sense serves as a secondary index for the data in the KVS. The LSM is needed for two purposes: lexicographic key order and 
\emph{
multi-version concurrency control (MVCC)}~\cite{WeikumVossen2001}. The former enables range queries, and the latter supports transactionally consistent snapshot reads that observe the values  of multiple keys at a common point in time.

Combining the KVS and LSM paradigms introduces a number of challenges, which we address in this work. Most importantly, we have to circumvent the indirection penalty of searching through the LSM when it is not required. 
To forgo this cost, we introduce the LSM bypass mechanism, which skips the LSM search altogether whenever the requested key has a single version. The main difficulty is to detect such situations quickly and to transition safely between the multiple-version and single-version scenarios. Much of the KV-Tandem algorithm is dedicated to handling this difficulty through a novel  use of Bloom filters and additional mechanisms.

Note that the LSM bypass approach greatly improves point get/put performance (by removing a level of indirection), but is disadvantageous for range scans, as the KVS does not support ordered iteration. This tradeoff is justified by our target workload, which uses range scans sparingly.

\subsection{XDP-Rocks} 

We materialize the KV-Tandem architecture in~\emph{XDP-Rocks} -- a full-fledged RocksDB-compliant storage engine, leveraging Pliops's XDP~\cite{Dayan-XDP} for the unordered KVS. XDP is hardware-accelerated to alleviate CPU bottlenecks related to key lookup, compression, GC, etc.; these features are orthogonal to KV-Tandem's design. The crucial aspect of the underlying KVS is the speed of random reads, which in XDP is close to the raw I/O speed thanks to the extremely lean memory footprint of its hashmap index that allows it to fit in DRAM even for large datasets.

XDP-Rocks is deployed by a number of commercial customers, which switched away from RocksDB to reduce costs and upgrade SLAs. A notable large-scale production deployment is at 
a hyperscaler video-sharing network. XDP-Rocks powers its 
internal distributed key-value database with Redis API~\cite{RedisAPI}, which supports dozens of consumer applications. Transitioning to XDP-Rocks allowed that operator to reduce its server fleet almost three-fold.

Building XDP-Rocks required solving a number of  nontrivial practical issues that arise in the KV-Tandem architecture. This includes shared storage management for LSM and KVS data, in-memory data caching, high-speed backup, and additional support for RocksDB's advanced APIs. Some of these challenges have been raised by industry users to enable transition from RocksDB. 

We provide an extensive evaluation of XDP-Rocks. We first compare it to BlobDB~\cite{BlobDB}, a state-of-the-art storage engine that implements WiscKey-style KV-separation in the RocksDB architecture.
We show that whereas BlobDB quickly runs out of space under write-heavy workloads due to its large SA, XDP-Rocks's SA is close to one. This allows us to support much larger datasets.
We then compare XDP-Rocks to RocksDB  in a broad set of RocksDB dbbench microbenchmarks~\cite{RocksDB} 
at production scale. XDP-Rocks exhibits higher average- and worst-case performance versus RocksDB in random writes, random reads, and combinations thereof. For example, we show a 3.5x throughput increase versus RocksDB under uniform random writes and 3.8x increase under mixed reads and writes. 
As expected, XDP-Rocks falls short of RocksDB in scan performance, but this can be mitigated through internal parallelism, to reduce the gap to only 20\% for a single scan.

We also evaluate XDP-Rocks at the system level, by using it instead of RocksDB as the storage engine of the popular Kvrocks distributed NoSQL database~\cite{KVRocks}. 
The experiments show high performance gains that correlate with the microbenchmarks -- e.g., a 20.5x throughput increase over a big dataset under a mixed read-write workload. 

\remove{
\idit{I wonder if there are any performance downsides in the evaluation that we should mention here. Niv mentions read-before-write. Indeed, we do try to read from the KVS even if the key is not in the LSM. Do we want to test that ? To revisit that ?

Another point Niv makes is that we're not comparing apples to apples since we use HW acceleration and the competition does not. Can we test with SW-only XDP?}
}
\subsection{Contributions and roadmap.}

This paper makes the following contributions:
\begin{itemize}
    \item We introduce a novel storage paradigm combining KVS and LSM  with a new approach to KV-separation; background on the existing paradigms we leverage is given in Section~\ref{sec:background}.
    \item 
    In Section~\ref{sec:kv-tandem}, we present the KV-Tandem ordered storage algorithm realizing this paradigm.
    \item We implement XDP-Rocks, an ordered storage engine based on KV-Tandem, in Section~\ref{sec:implementation}.
    \item In Section~\ref{sec:results} we evaluate XDP-Rocks through microbenchmarks and system-level tests. We compare it with both RocksDB (performance baseline) and XDP (performance upper-bound). 
    \item Section~\ref{sec:discussion} discusses KV-Tandem's advantages and limitations and outlines future work.
\end{itemize}


\section{Background and Previous Work}
\label{sec:background}

We now provide some essential background for this work. 
Our goal is to implement the API of a rich ordered storage engine, e.g., RocksDB~\cite{RocksDB} described in Section~\ref{sec:API}.
Our design extends the LSM-tree paradigm, which we discuss in Section~\ref{sec:lsm}. It shares some ideas with prior work on KV-separation but also differs from it profoundly, as discussed in Section~\ref{sec:kv-sep}. 

\subsection{Ordered storage engine API}
\label{sec:API}

Storage engines~\cite{LevelDB, RocksDB, Pebble, WiredTiger} organize KV pairs in containers called \emph{databases}~\cite{RocksDB} (also known as tablets~\cite{ChangDGHWBCFG08} and collections~\cite{WiredTiger}). A database is a unit of storage management (space allocation, security, replication, etc.) and a gateway to all KV APIs. 

Most storage engines provide  the following basic APIs:
random (point) read --  \code{get(k)}, range read -- \code{iterate(from\_k,to\_k)}, and random-write -- \code{put(k,v)} and \code{delete(k)}. The \code{iterate} API  allows traversing either the keys or the KV pairs within a range, in either ascending or descending order. The put API stands for either an insert of a new KV pair or a value update for an existing key. (Additional APIs can be provided, but for clarity, we focus on these).

It is worth noting that puts and deletes do not guarantee immediate durability (the update might not persist through crash failures) because they typically return after writing in memory without waiting for a disk flush.

Common ordered storage engines (e.g.,~\cite{RocksDB, WiredTiger}) also provide higher-level transaction APIs that facilitate ACID semantics for multiple reads and writes. For performance reasons they typically favor the snapshot isolation consistency model, which in particular guarantees that all reads pertain to the same point-in-time view of the dataset, called \emph{snapshot}. They thus offer APIs for snapshot creation and for reading in it  -- \code{get@sn(k,sn)} and  \code{iterate@sn(from\_k,to\_k,sn)}. 

Storage engines also provide more eclectic APIs, e.g., range delete, checkpoint (persistent snapshot), compaction filter (auto-expiration), and others in RocksDB~\cite{RocksDB}. We do not focus on these functions here unless supporting them projects on KV-Tandem.

\subsection{LSM trees}
\label{sec:lsm}


KV-Tandem employs the popular LSM data structure~\cite{LSM96}, which is  used by many production ordered storage engines~\cite{LevelDB, Pebble, RocksDB, Speedb}. This section is an essential summary of LSM structure for understanding our KV-Tandem solution.

LSMs optimize write I/O by transforming small random writes to I/O-friendly big sequential writes, as follows. All incoming updates are staged in a sorted memory buffer called \emph{memtable}, typically 64-128 MB big~\cite{RocksDB}. Whenever memtable reaches capacity, it is flushed into storage as a sorted  file called a \emph{sorted string table} (SST). The SST file becomes immutable and searchable at the end of flush. SST files are later merged through a background \emph{compaction} process, which also removes obsolete (over-written or deleted) data.

Level capacities grow exponentially.
The topmost level, L0, is filled by flushes. Every lower level consists of a collection (also called \emph{run}) of SST files with disjoint ranges. Whenever a level reaches capacity, some SST 
file is picked to be merge-sorted with intersecting SST
files at the next (lower and larger) level(s). 

The LSM components (memtable and SST files) store $\langle$\code{key}, \code{sn}, \code{value}$\rangle$ triples, each capturing a distinct \emph{version} of \code{key}. The \code{key}\ and \code{value}\ are application-provided, whereas \code{sn} is a logical timestamp 
(sequence number) generated by a global logical clock (counter) that is promoted upon every update. 
A \emph{snapshot handle} \code{S} is the \code{sn} at the time of that snapshot's creation. A simple read of key \code{k} returns its last written version. A read from snapshot \code{S} returns the value of the latest version of \code{k}\ whose sequence number precedes \code{S}. 

Each SST file is sorted by $\langle$\code{key}, \code{sn}$\rangle$ to provide logarithmic-time binary search for random reads and sequential access  in range reads. LSMs typically maintain a range-to-block \emph{index} per SST file in order to restrict key searches to one block per SST file.  The index  is concise and is cached in RAM with high priority.

Since the ranges of SST files in different levels overlap, gets need to search all LSM levels, from 
new to old.
To mitigate the RA cost, LSMs maintain, per SST file, a \emph{Bloom filter}~\cite{ChangDGHWBCFG08} -- a succinct statistical summary of the keys in that file. Bloom filters have no false negatives 
and low-probability false positives (infrequently return true for a non-existing key). Thus, the actual file search (and I/O) can be skipped in most cases when the key is not present in the file. 
Bloom filters, like SST indexes, are cached with high priority. 
Note that the Bloom filter optimization only applies to random reads. Range-read queries merge the range content from multiple SST files.

To facilitate data durability between the insertion into memtable and the flush, every update is 
persisted to a \emph{write-ahead log} (\emph{WAL})
within a short time of its occurance.  
The WAL is not searchable; its primary use is during recovery from crash failures. Once memtable is persisted in the form of an SST file, the WAL entries pertaining to that data  may be recycled.

Deletions are implemented by storing 
a special $\langle\code{k}, \code{sn}, \bot\rangle$  version, called \emph{tombstone}, which designates a deletion of key \code{k} at time \code{sn}. A $\langle\code{from\_k}, \code{to\_k}, \code{sn}, \bot\rangle$ tombstone designates a deletion of a whole range. A get that  retrieves a tombstone, either  single-key or range, returns no result. 
Compactions that encounter tombstones remove the corresponding deleted entries from the LSM.
A compaction that spans the lowest LSM level also eliminates the tombstone itself (as long as no earlier version survives); at higher levels, the tombstone is kept in the SST file.

\remove{

The number of disparate sorted ranges (\emph{runs}) in an LSM affects the RA of all query types. Every flush creates a new run (SST file). LSM trees bound the number of runs (and consequently, the worst-case RA) through a background process named \emph{compaction}, which merges small runs into bigger ones.  One other function of compaction is eliminating the old key versions that are no more accessible to queries, thereby bounding the SA too. Note that historical versions may not be removed if they may be required by some active snapshot read. Once a compaction completes, a new run (one or more immutable SSTs) becomes part of the LSM, and the input SSTs are discarded.}

\remove{LSMs shape there storage as multi-level tree through flushes and compactions. The sorted runs are organized in \emph{levels} (one or more per level). The topmost level, named L0, is filled by flushes. A compaction is triggered when some level fills up. It may merge any number of contiguous levels into the next-lower level (e.g., levels L3--L5 into level L6). Level capacities grow exponentially, thereby most of the data sinks to the bottommost levels. Figure~\ref{fig:LSM} depicts the LSM geometry.}

\remove{
\begin{figure}
    \centering
    \caption{LSM Tree structure.}
    \label{fig:LSM}
\end{figure}
}

\subsection{LSM performance and KV-separation}
\label{sec:kv-sep}

While compactions reduce the RA and SA metrics, they also increase the WA, as the data is copied across levels multiple times. 
For large datasets, the WA can exceed 50x, leading the system to utilize less than 2\% of the storage device's bandwidth~\cite{LuoCarey}. Furthermore, compactions 
cause
unpredictable performance fluctuations~\cite{GiladBBGHKMS20, LuoC19a}. A host of research studied different compaction policies to navigate tradeoffs in the LSM performance space (see  survey in~\cite{AthanassoulisIS23}), but the tradeoffs between the metrics are inherent to the LSM approach. 

\paragraph{Key-value separation}
To overcome this conundrum, the seminal WiscKey paper~\cite{wiscKey} suggested to store data values separately from the LSM to avoid repeatedly compacting them and thereby reduce WA. WiscKey stores the values in external files called value-logs (\emph{v-logs}) and uses an LSM to keep  $\langle$\code{key}, \code{sn}, \code{loc}$\rangle$ triples, where 
\code{loc} is the value location (v-log id and offset). Consequently, only the keys and their concise metadata get re-written by compactions, rather than the whole dataset. In many KV-storage applications, the values are an order of magnitude bigger than the keys~\cite{CaoDVD20, CooperSTRS10}, so this approach drastically reduces WA. 

\paragraph{Indirection overheads}
The KV-separated LSM is essentially a secondary index into the v-log storage. Every read starts from searching the LSM for the value reference and then retrieves the value from a v-log file. This two-hop access scheme increases RA (affecting read latency and throughput)  unless the \emph{whole} (albeit reduced) LSM tree fits into memory. 
In reality, a system might not have that much RAM --   some systems have a RAM-to-storage ratio of 2\% and below~\cite{ClickhouseSizing},  meaning that the LSM-tree cannot be fully cached. KV-Tandem eliminates this constraint through LSM bypass, which skips the LSM search altogether. 

\paragraph{GC overheads}
Another challenge introduced by KV-separation is garbage collecting  obsolete values from the value log. Note that LSM stores \emph{physical} v-log addresses, so value relocation by GC incurs I/O overhead. 
WiscKey's GC accesses the LSM twice for each relocated value, first to query whether the value is still valid and then to update the address mapping.
This mechanism manifests in significant write rate loss~\cite{wiscKey}.
HashKV~\cite{Chan18} improves upon WiscKey's GC by partitioning the v-log by keys, which allows them to check validity within a v-log partition without an LSM query. However, it increases the SA due to fragmentation. Production KV-separated LSMs, e.g., BlobDB~\cite{BlobDB} and TerarkDB~\cite{Terarkdb}, 
have lazy GC: they track value invalidations at compaction time and discard a v-log file as a whole once all its values become invalid. This design implies unbounded SA under heavy writes (as we show in Section~\ref{sec:results}), which is impractical in many cases. KV-Tandem's use of a KVS decouples the values' GC from the LSM and imposes no constraints on physical GC.

\paragraph{Handling small values}
KV-separation is not beneficial for small values, as it fails to improve WA significantly while putting extra pressure on RA and SA. Practical systems take a hybrid approach, storing large values in v-logs while embedding the rest in the LSM. For example, BlobDB~\cite{BlobDB} recommends to apply KV-separation only to blobs 1KB and above. In XDP-Rocks, we also support this hybrid approach, namely, embed small values in the LSM. Because XDP-Rocks uses an optimized KVS with very low per-value overhead, it reduces the separation threshold to 128--256B.

\remove{
KV-separated LSMs are also inherently inferior to the traditional design in range-read performance, since their v-log storage does not preserve the order of keys. They retrieve every value separately, thus not only failing to exploit sequential  I/O but also introducing RA to the random I/O, especially for sub-4-KB values. A host of research suggested hybrid designs to mitigate that gap through limited locality of storage (DiffKV~\cite{LiLiu21}), moving the values in and out the LSM (Parallax~\cite{XanthakisSBPB21}), etc. KV-Tandem is complementary to that work, since it applies to all values stored outside the LSM. 
}


\section{KV-Tandem: Marrying KVS and LSM}
\label{sec:kv-tandem}

We now present KV-Tandem, a new KV storage paradigm that combines KVS with an LSM index and bypasses the latter in the common case. KV-Tandem stores data in the KVS in two distinct key-encoding modes, \emph{direct} and \emph{versioned}, which we introduce in  Section~\ref{ssec:basic}.  Section~\ref{sec:read-write-alg} then presents the algorithm's basic read and  write  paths.
The combination of independent KVS and LSM engines introduces challenges in recovering from crashes; we address them in Section~\ref{sec:recovery}.

\subsection{Storage modes}
\label{ssec:basic}

Our approach is founded on the observation that most keys in an ordered KV storage engine are associated with a single value (version). Multiple versions need to be kept only if a key is overwritten in the course of an ongoing snapshot, whereas
common workloads of ordered KV storage engines, e.g., Redis-on-Flash~\cite{RoF}, apply snapshots only in specific scenarios such as backup.

To optimize for the common single-version case while maintaining the ability to support multi-versioning, we use two distinct key encoding schemes in KVS (storage modes). Keys that have a single version are (typically) stored in \emph{direct} mode in the KVS, and can be accessed without the LSM. Other keys are stored in \emph{versioned} mode, and their access requires indirection via the LSM. These modes are transparent to the KVS, which handles all keys identically. 

\paragraph{Direct storage}

In direct mode, we keep KV-pairs in the KVS directly under their application-provided keys. Every value stored in this mode is concatenated with its LSM-generated sequence number (sn), so that snapshot queries can verify whether this value pertains to their snapshot.

Note that even if all keys are stored in direct mode (i.e., each has a single version), the KVS is insufficient by itself because it does not support  range scans. To this end, we use an LSM  as an ordered secondary index for the data in the KVS. 

Consider, for a moment, the idealized scenario whereby keys always have at most one version. Here, the implementation of KV-Tandem's put and get operations could have been straightforward, as shown in Algorithm~\ref{alg:ideal}: A \code{get(k)}\ searches for \code{k} in the LSM memtable first and then in the KVS. A \code{put(k,v)}\ inserts the KV-pair into the memtable with the LSM's current timestamp (sn) (not shown). 
When  memtable becomes full, a \code{flush} stores its data both in the KVS and in the LSM.
Flush iterates over the memtable, which is sorted by key. Redundant entries are filtered out, and \code{flushEntry} is called for each of the surviving 
$\langle$key, sn, val$\rangle$ 
triples.  If multiple entries of the same key are flushed, they are flushed in decreasing \code{sn} order, namely, from newest to oldest. Every such pair is stored in the KVS via a blind write, potentially overwriting the previous value.
The key and version are also stored in the new L0 SST file. 

When flush completes, the new L0 SST file is connected to the LSM tree, the memtable is emptied, and the WAL is truncated. 

\begin{algorithm}[htb]
\begin{multicols}{2}
    \begin{algorithmic}
    \Procedure{get}{k} 
          \State ret $\leftarrow$ memtable.get(k)
          \If{ret $\not=$ null} return ret \EndIf
          \State return KVS.get(k) \Comment{strip sn}
    \EndProcedure
    \Procedure{flushEntry}{F, k, sn, val} 
    \Statex \Comment{flush  memtable entry  to SST file F}
        \State KVS.put(k, sn$\mathbin\Vert$val) \label{line:put-simple}
        \State F.add(k, sn, \code{null}) \Comment{empty value}
            \label{line:sst-add-simple}
    \EndProcedure
    \end{algorithmic}
    \end{multicols}
    \caption{Idealized KV-Tandem with direct storage only (single version per key).}
    \label{alg:ideal}
\end{algorithm}

Note that the LSM  might keep multiple versions of the same key in different SST files, even if they are not needed for any snapshots. This occurs when a value is overwritten while its previous (now obsolete) version(s) reside(s) in SST files pertaining to lower (older) levels.
WiscKey-style key-value separationv~\cite{wiscKey} would keep all these versions in the log storage until 
a GC process is triggered, which in BlobDB~\cite{BlobDB} and TerarkDB~\cite{Terarkdb} can take indefinitely long.
In contrast, the KVS  has its own built-in GC, allowing the overwritten KV pairs to get garbage-collected fast.  

Direct storage is very effective for point queries, which always retrieve the latest written value. But for supporting snapshots (as needed for ACID transactions), we might need to keep multiple versions of a given key, as we discuss next.

\paragraph{Versioned storage}

When a snapshot is created, it logically ``freezes'' a copy of the database at the current point in logical time (sequence number). 
The list of the active snapshots' sn's is kept in-memory (as in other LSM implementations~\cite{RocksDB}).
As long as the snapshot is live, new writes must refrain from overwriting the latest values written before the snapshot's timestamp. Hence, we cannot write new values of existing keys in direct mode. 

Instead, when we write a new value during a live snapshot, we store it in the KVS in versioned mode, namely, indexed by its \code{key-sn} pair. 
(In this mode, there is no need to concatenate the version number to the value, as it is part of the key.)
Thus, the KVS  simultaneously contains values indexed in direct mode (by their keys only) and others in versioned mode. To distinguish between the two cases, we add a bit to the LSM entry: 
LSM entries become $\langle$key, sn, vm$\rangle$ triples, where \code{vm} indicates if version \code{sn} of \code{key} is stored in versioned mode in KVS.

Writing in versioned mode can create a situation where there is an old value of a key  in direct mode and a newer one in versioned mode. 
Get queries (which do not pertain to a snapshot) must be able to determine which version is the latest. We achieve this by
ensuring that direct values are always older than all versioned values. To this end, every update of a key that already has versioned values in the storage is done in versioned mode, regardless of active snapshots. In this way, we guarantee the following invariant:

\begin{invariant}[Direct is older] 
\begin{description}
\item[\newline]
\item[KVS] 
If a key occurs in KVS in both direct and versioned modes, then the value in direct mode is older than all versioned values.
\item[LSM]
If a key occurs in LSM in both direct and versioned modes, the values in direct mode are indexed in lower or equal LSM levels than all versioned values.
\end{description}
\label{invariant:direct}
\end{invariant}

Note that a key can occur in KVS
at most once in direct mode 
and multiple times in versioned mode  (paired with different sn's). In the latter case, the version in the highest LSM level is valid, and the lower ones might be required for snapshots reading ``in the past'' or might be obsolete.

Given Invariant~\ref{invariant:direct},  queries can always begin their search top-down in the LSM and fall back on direct search in the KVS in case the LSM search yields no  results. But this means that every \code{get} must now go through the LSM. 
Moreover, to ensure Invariant~\ref{invariant:direct},  \code{flushEntry} must check if the written key already appears in versioned-mode in the LSM.
We next explain how 
our LSM bypass mechanism avoids the cost of a full LSM search in  common cases.

\subsection{KV-Tandem read and write paths}
\label{sec:read-write-alg}

We first explain the LSM bypass in Section~\ref{sec:lsm-bypass}, and then delve into the algorithm's details, covering gets and flushes in Section~\ref{ssec:flush}, compactions in Section~\ref{sec:compaction}, and
snapshots and range operations in Section~\ref{sec:snapshot-get}.

\subsubsection{LSM bypass}
\label{sec:lsm-bypass}

An LSM search may involve many (potentially tens of) files. 
This can be very  slow, particularly if the LSM is too big to fit 
in its entirety in RAM, and defeats the whole purpose of using direct access. To avoid this cost, we design a novel \emph{LSM bypass} mechanism that allows queries to quickly discover that a key is in  direct storage, which we assume happens most of the time. 

To support the LSM bypass, we repurpose the LSM's Bloom filters -- instead of indicating whether a key exists as in a standard LSM SST file, the Bloom filter now indicates whether the key is present and \emph{is stored in the KVS in versioned mode}. 
If key is present in the SST but has no versioned mode in KVS, the Bloom filter returns false, and the SST search is skipped. Because a Bloom filter may return false positives, a query that does search the SST file must still check the \code{vm}\ bit in the appropriate SST entry. 


Importantly, the Bloom filters are very compact ($\sim$10 bits per key) in traditional LSMs~\cite{DayanAI17}, and  more so in KV-Tandem, where versioned entries are a small fraction. As long as the Bloom filters reside in RAM, 
this mechanism eliminates the bulk of the I/O cost of the LSM search. A Bloom filter search does require some computation cost for computing hash functions, but since the LSM uses the same hash functions in all of its SST files, the hash keys can be computed in a pre-process before the LSM search and so the overhead for multiple Bloom filter searches in negligible.

\begin{algorithm}[thb]
\begin{algorithmic}[1]
\Procedure{get}{k}
\label{line:start-get}
    \State ret $\leftarrow$ memtable.get(k)
    \If{ ret $\not=$ null} return ret \EndIf 
    \For{each SST file F included in the LSM search (new to old)}    \label{line:SST-get}
        \If{F.inBloom(k)} \Comment key might be in versioned mode
             \State entry $\leftarrow$ F.searchLatest(k) \Comment{find entry with highest \code{sn} for \code{k} in F }
                    \label{line:LSM-search}
            \If{entry $=$ null}  \label{line:LSM-no-entry}
                continue
                \Comment{false positive, ignore}
            \EndIf               
                \If{$\neg$entry.vm} break \label{line:check-vm}
                    \Comment key found in direct mode, exit search loop
                \EndIf   
                \State ret $\leftarrow$                      KVS.get($\langle$entry.k, entry.sn$\rangle$)
                \Comment{versioned-mode KVS search}
                \label{line:KVS-search}
                \If{ret = null} break
                    \label{line:fallback}
                    \Comment versioned key concurrently removed by rename
                \EndIf
                \State return ret
                \label{line:end-versioned-search}
         \EndIf   
   \EndFor
\State return KVS.get(k) \Comment{direct mode search, strip sn}
\label{line:end-get}
\EndProcedure

   \Statex

\Procedure{flushEntry}{newF, k, sn, v} 
        \Comment flush  memtable entry  to SST file newF
    \label{line:start-flush}
     \If {isDirectModeSafe(k, sn, $0$)} 
            \State KVS.put(k, sn$\mathbin\Vert$v)
            \label{line:kvs-add-false}
                                    \Comment  direct mode
            \State newF.add(k, sn, false) \Comment{does not add k to Bloom filter} \label{line:add-false} 
    \Else 
            \State KVS.put($\langle$k, sn$\rangle$, v)
                                    \Comment  versioned mode
            \label{line:kvs-add-true}
            \State newF.add(k, sn, true) \Comment{adds k to Bloom filter}     \label{line:add-true}
    \EndIf

    \Statex 
    
    \Procedure{isDirectModeSafe}{k, sn, lvl}

        \If{there is an active snapshot earlier than sn}
            \label{line:active-snapshot}
            return false
        \EndIf
        
        \For{each SST file F in LSM search for k below level lvl} 
            \label{line:is-versioned}
            \If{F.inBloom(k)} 
                return false \Comment key (maybe) found in versioned mode
            \EndIf  
        \EndFor
        \State return true 
    \EndProcedure

    \label{line:end-flush}
    \EndProcedure

    \algstore{algo}
    \end{algorithmic}
    \caption{KV-Tandem with LSM bypass.}
    \label{alg:simple}
\end{algorithm}

\subsubsection{Get and flush}
\label{ssec:flush}

The basic KV-Tandem algorithm is shown in Algorithm~\ref{alg:simple}.
The \code{get}\ operation (line~\ref{line:start-get}--\ref{line:end-get}) searches for the latest
version of the given key. 
The traversal order (in lines~\ref{line:SST-get} and also~\ref{line:is-versioned}) is  as in a regular LSM search, top to bottom. In each layer, the SST file covering the appropriate key range is searched; in L0 all files are searched.

Get first checks the file's Bloom filter. Whenever a Bloom filter returns false, there is no need to fetch the SST file's data. 
Otherwise, we retrieve the latest version of the key from the SST file in line~\ref{line:LSM-search}. If it is not found, 
(line~\ref{line:LSM-no-entry}), we know that the Bloom filter returned a false positive and continue the search. 
If the key is found but the SST entry indicates that it is in direct mode (line~\ref{line:check-vm}), this also means the Bloom filter returned a false positive, but in this case, a direct version exists, so we stop the search and revert to direct mode.   
If all SST searches return false, it means that the key does not occur in versioned mode, and a direct mode \code{get} is performed (line~\ref{line:end-get}).

As we explain below, the compaction process might delete a versioned entry of a key from both the KVS and the LSM (as part of a processes called \emph{renaming}), and this does not occur atomically.
Therefore, it is possible for \code{get}\ to find a key-version pair in the LSM in line~\ref{line:LSM-search} and then fail to find it in the KVS in line~\ref{line:KVS-search} (because it had been already deleted from the latter but not yet from the former). In this case,  
we also fall back on a direct-mode KVS search.

The flush procedure builds a new SST file, including the Bloom filter, via a sequence of calls to \code{flushEntry} 
(lines~\ref{line:start-flush}--\ref{line:end-flush}). To avoid races between checking for the existence of versioned values and writing them, we flush one memtable at a time (oldest first), as is often the practice in LSMs~\cite{RocksDB}.

Flush also writes to KVS the KV-pairs pertaining to the  SST entries in either direct or versioned mode, as determined by the \code{isDirectModeSafe} predicate. 
(The \code{lvl} parameter in the predicate is used to restrict the search in compactions, as described below; in flush \code{lvl} is always $0$, 
so the entire LSM tree is searched). 
In  case versioned mode is needed, 
the key is inserted into the Bloom filter (line~\ref{line:add-true}).
We revert to versioned mode in two cases: 
\begin{enumerate}
    \item The update is spanned by an active snapshot (line~\ref{line:active-snapshot}).  
    \item The key has older values in versioned mode (line~\ref{line:is-versioned}). Here, too, only Bloom filters are checked and we forgo a full-fledged LSM search.
\end{enumerate}

Since line~\ref{line:is-versioned} only checks
existing SST files, it does not account for the scenario that different versions of the same key are flushed into one output SST file. 
Fortunately, this is addressed by the first condition: two versions of the same key are retained only if the newer of the two is spanned by an active snapshot. Thus, 
\code{isDirectModeSafe} returns false for this (newer) version in line~\ref{line:active-snapshot}.

\remove{
since the versions are written in decreasing \code{sn} order, whenever the first condition holds for an earlier version than one that had already been flushed to the same file, the condition must have been evaluated to true for the previously flushed newer version as well. In fact, the older version is only retained in the output file if it is needed for an active snapshot, meaning that the newer version satisfies condition 1.
}

Whenever all Bloom filters are cached in RAM, \code{isDirectModeSafe} does not perform any I/O. 
The price we pay for this is that \code{isDirectModeSafe}
may (infrequently) produce false negatives, i.e., imply versioned mode where direct mode would have been safe.
This does not compromise safety, because writing new values in versioned mode preserves Invariant~\ref{invariant:direct}.

\remove{No disadvantage at checking tens of Bloom filters instead of a complex common data structure indicating storage mode for all. Sub-microsecond latency per Bloom versus 30-50 microsecond per I/O.}

\subsubsection{Compaction}
\label{sec:compaction}

The LSM's \emph{compaction} process~\cite{LuoCarey} merge-sorts collections of SST files in the background while garbage-collecting obsolete data. A compaction creates one or more new (output) files to replace multiple existing (input) files (in a contiguous sequence of layers), which are then discarded.  
%
It iterates over the LSM entries in the input files and decides which of them to keep.
An entry is kept in two cases:
\begin{enumerate}
    \item The entry holds the highest (latest written) version of its key.
    \item The entry holds an overwritten version that is still required for an active snapshot. Namely, this is the last version written before the snapshot time.
\end{enumerate}


Because SST files are sorted, these conditions can be checked locally during the  merge-sort. Upon determining that an entry is obsolete (i.e., does not satisfy either of the above conditions), it is removed from KVS via the \code{compactionDelete}\ procedure in lines~\ref{line:start-delete}--\ref{line:end-delete} of Algorithm~\ref{alg:compact}. 
Otherwise, the entry is inserted into the new SST file via the \code{compactionWrite}\ procedure in lines~\ref{line:start-compact}--\ref{line:end-compact}. Tombstones are handled the same way, i.e., they render obsolete earlier versions of the key that are not needed for snapshots.


\begin{algorithm}[tbh]
    \begin{algorithmic}[1]
    \algrestore{algo}


    \Procedure{compactionDelete}{k, sn, vm} 
    \Statex 
    \Comment remove from KVS an obsolete version found in an  LSM entry in a compaction input file 
    \label{line:start-delete}
    \If{vm} \Comment{versioned entry}     
        \State KVS.delete($\langle$k, sn$\rangle$) 
        \label{line:versioned-delete}
    \Else \Comment{direct mode entry}
        \State KVS.delete(k) 
    \EndIf    
    \label{line:end-delete}
    \EndProcedure

    \Procedure{compactionWrite}{newF, k, sn, vm} 
    \Statex
        \Comment add an LSM entry from a compaction's input file to a new SST file \code{newF}\ (after filtering)
    \label{line:start-compact}
    \If{$\neg$vm} \Comment{direct mode entry remains direct}
            \State newF.add(k, sn, false) \Comment{does not add k to Bloom filter}
    \ElsIf{$\neg$isDirectModeSafe(k, newF.level)} 
        \Comment{cannot rename}
        \State newF.add(k, sn, true) \Comment{adds k to Bloom filter}
    \Else \Comment{rename}    
    \label{line:start-switch}
        \State v $\leftarrow$ KVS.get($\langle$k, sn$\rangle$)
        \State KVS.put(k, sn$\mathbin\Vert$v) 
        \State KVS.delete($\langle$k, sn$\rangle$)
        \State newF.add(k, sn, false) \Comment{does not add k to Bloom filter}
    \EndIf
    \label{line:end-switch}
    \label{line:end-compact}
    \EndProcedure
    \end{algorithmic}
    \caption{KV-Tandem compaction.}
    \label{alg:compact}
\end{algorithm}

In the scheme described thus far, once a key has a value in versioned mode, all its future versions are also stored in versioned mode (to maintain Invariant~\ref{invariant:direct}). This does not undermine correctness -- indeed, it is safe to store all keys in versioned mode -- but may have performance implications, particularly for long-lived frequently written keys that are seldom required for snapshots. 

To address this, we allow compactions to \emph{rename} values in the KVS, namely 
switch them  from versioned to direct mode. A rename reads the old (versioned) KV-pair, writes a new (direct) KV-pair with the same value, and finally deletes the old KV-pair from KVS (lines~\ref{line:start-switch}--\ref{line:end-switch}). 
Rename occurs whenever it is safe to use direct mode for the key, which is checked by the \code{isDirectModeSafe} predicate of Algorithm~\ref{alg:simple}.  In this case, we only check SST levels below the new output file, as needed to satisfy Invariant~\ref{invariant:direct}.  

\remove{
We next explain how we discover that a rename is safe. 


Consider first the lowest (oldest) LSM level. This level holds most of the keys. 
Now consider a key that has a single version in a new bottom-level SST file. Recall that this is the common case, since  multiple versions of the key are included in the output file only in case old versions are needed for active snapshots. 
Because there are no lower levels, Invariant~\ref{invariant:direct} trivially holds for such keys, regardless of their storage mode. Therefore, every compaction into the bottom level can write in direct mode 
all keys for which it does not retain multiple versions.

We can extend this logic to any LSM entry that is the lowest versioned one \inred{in the tree}.

Consider a \code{key-sn} pair written to an output SST file in some level \code{lvl}. 
If there is no active snapshot earlier than \code{sn}, 
\inred{then the key is not retained with multiple versions in the output SST file. Therefore, if }
the key has no versions in levels below \code{lvl}, Invariant~\ref{invariant:direct} holds, and it is safe to keep it in direct mode.  
This is checked in \code{isDirectModeSafe} (lines~\ref{line:start-lowerVersion}--\ref{line:end-lowerVersion}), which generalizes the same predicate in Algorithm~\ref{alg:simple}. 
}

Notice that rename is uni-directional: written values never move back from direct mode to versioned mode. Moreover,  direct-mode writes are always blind -- i.e., they are oblivious to whether they overwrite an obsolete direct KV-pair or create a new one.
Thus, the impact of rename on write amplification is small -- at most one extra write per put.


\subsubsection{Snapshot reads}
\label{sec:snapshot-get}

As in a standard  LSM, creating a snapshot returns the current logical time and adds it to an in-memory list of active snapshots. A snapshot is released by removing it from the list. Note that snapshots are ephemeral by nature, hence there is no need to persist them. 

Once a snapshot has been created, it is possible to read in it.
The implementation of \code{get@sn} is similar to that of \code{get}, with the addition of version matching in the two read paths, as follows:
\begin{description}
    \item[Versioned mode:] 
    The \code{searchLatest(k)} call in line~\ref{line:LSM-search} is replaced by a \code{searchLatestBefore(k,sn)} call, which returns the highest version \emph{smaller than} \code{sn} of \code{k} in F. If there is no such version, it returns null. From then until line~\ref{line:end-versioned-search}, the code is the same.
    \item[Direct mode:]
    Before returning the value read in line~\ref{line:end-get}, its version is compared to the snapshot version. If it is not smaller than the snapshot timestamp, we deduce that there is no version earlier than the snapshot (because the direct version is always the oldest), hence \code{get@sn} returns null.
\end{description}

The \code{iterate} API is implemented by taking a snapshot,  calling \code{iterate@sn} in it, and then disposing it. \code{iterate@sn} merge-sorts the relevant SST files at all LSM levels to retrieve 
all versions between the first key and the last key, and selects for each key the latest version before \code{sn} through \code{searchLatestBefore}. For each selected version, it looks up its value in KVS, using the \code{vm} bit to discern the access mode. LSM bypass does not apply to iterators. 

\remove{
Atomic range deletions are represented by special range tombstones holding $\langle$minkey, maxkey, sn$\rangle$ triples. These tombstones are maintained in a global in-memory interval tree. Queries treat range tombstones similarly to snapshots: 
A \code{get} query always checks this data structure, and if the requested key is in a deleted range, the range tombstone's \code{sn} is compared to the latest stored version of the key in order to determine whether to return the latest version (if it is newer than the tombstone) or null. A \code{put} of a key that is covered by a range delete is always done in versioned mode
(here the range delete is interpreted as an active snapshot). }

\subsection{Recovery from crash failures}
\label{sec:recovery}

LSM storage is modified by two background processes -- flush and compaction. 
In traditional LSMs, a failure during either of them has no effect, as the partially written SST files are discarded upon recovery. In KV-Tandem, recovery is trickier because  flush and  compaction also update the KVS, and these updates do persist after the crash. We now describe how KV-Tandem handles these post-recovery effects. 

\paragraph{Flush}
As in other LSMs, recovery from crash failures is facilitated by the WAL, which tracks all updates that have not yet been flushed into the LSM. 
Upon recovery from a crash, the WAL contains the updates that happened after the latest flushed sequence number. The recovery procedure promotes the logical clock beyond the pre-crash value, and then replays the WAL. 


In KV-Tandem post-recovery, 
the KVS contains some subset of the WAL entries. 
Unfortunately, these entries do not necessarily correspond to  ones 
that will be added by replaying the WAL,  because updates might be done differently the second time. To see why, consider a value that has been stored by a pre-crash flush in versioned mode, either because of an ongoing snapshot or because the key is versioned in some older SST. Both situations are ephemeral -- snapshots do not survive system restarts, and versioned entries in lower-level SSTs might have been removed by compactions that overlapped the flush and did complete successfully before the crash. If none of the above holds, the post-recovery flush stores the value in direct mode. Otherwise, the new flush writes the value in versioned mode, but uses a new (post-crash) version number for the new update.
In both cases, the KVS still contains an \emph{orphaned} version that was inserted during the pre-crash flush and is not referenced by any SST.  If left un-handled, this poses a permanent storage space leak.

To address that, we extend the recovery with an \emph{undo} step, which removes the orphaned values from the KVS. The WAL contains the original \code{sn} assigned to the entry before the crash. If a partial flush had written that key in versioned mode, it had used this version. We thus preemptively delete the corresponding versioned entry from the KVS (if it does not exist, the delete is void). Tombstones are skipped because they do not create values in the KVS.

Note that there is no need to undo direct-mode updates. Because there are no active snapshots during recovery and compactions can only eliminate versioned values, all updates done in direct mode before the crash will be re-done in direct mode in the redo phase. Thus, any direct-mode values added before the crash will be simply overwritten upon recovery.

\paragraph{Compaction}
Unlike flushes, compactions are not replayed, since they process the data that is already part of the LSM tree. They resume lazily after recovery. There is no guarantee that the post-recovery compactions span the same sets of SSTs as pre-crash.

Partially executed compactions may affect the KVS in two ways -- renaming (Algorithm~\ref{alg:compact}, lines~\ref{line:start-switch}--\ref{line:end-switch}) and deletion of removed or overwritten keys. Fortunately, deletions are idempotent. If a post-crash compaction tries to delete a value that had already been deleted or renamed, the KVS delete does nothing. 


However, renaming leaves a dangling pointer in the old SST file and an orphaned direct-mode value in KVS. Note that such dangling pointers also temporarily arise in the course of compactions in non-crash scenarios; KV-Tandem addresses them by falling back on direct-mode search in case the versioned value is not found (line~\ref{line:fallback}). The dangling pointer is ephemeral -- eventually, the old SST file will be recycled by some future compaction, and if the key will survive the compaction, it will be renamed to direct mode. 
If the key does not survive the compaction, however, the orphaned version must be removed to avoid a storage space leak. To this end, we modify \code{compactionDelete} so that whenever a compaction that spans the bottommost LSM level removes a versioned-mode entry of a key
(line~\ref{line:versioned-delete}), it proactively removes the direct version of the same key as well. By  
Invariant~\ref{invariant:direct}, this version is necessarily obsolete so removing it is safe.

\section{XDP-Rocks}
\label{sec:implementation}

XDP-Rocks is a RocksDB-compatible storage engine that implements the KV-Tandem design atop XDP~\cite{Dayan-XDP}, a hardware-accelerated KVS technology, which we overview in Section~\ref{sec:xdp}. 
XDP-Rocks addresses multiple practical challenges that are not addressed by KV-Tandem per se but are required in production; these are described in Section~\ref{sec:xdprocks}.

\subsection{XDP background}
\label{sec:xdp}

XDP~\cite{Dayan-XDP} is a storage SDK that provides a KVS API~\cite{KVSSD}. The host library is a thin software layer on top of a storage processor PCIe card. XDP communicates with the storage directly, bypassing the filesystem, using proprietary data layout and index. 
The storage processor manages either directly-attached or over-fabric NVMe SSDs.
It offloads index update, lookup, and data compression to FPGA circuitry. XDP also provides transparent RAID5 reliability. 

The XDP software model supports variable-size logical shards, or \emph{databases}, to partition the storage space. Database creation and deletion are instant atomic operations. KVS operations (get, put, and delete) are provided at the database level. Additionally, XDP provides a full database unordered scan API for analytics and replication. 
XDP incorporates efficient GC (with just 7\% overprovisioning), which 
facilitates database deletion  with zero extra I/O.

The XDP KVS is essentially a hashtable atop log-structured storage. Its spirit is similar to Bitcask~\cite{Bitcask} and Aerospike~\cite{SrinivasanBCSGI16}, but it scales better than them. 
The memory overhead is between 2.5 B and 3.5 B per key, depending on the expected value size -- an order-of-magnitude less than the closest competition~\cite{SrinivasanBCSGI16}. This allows maintaining the whole index in DRAM, i.e., the key-to-address mapping incurs no extra I/O. XDP has no additional data cache in host DRAM; this is left to the application. 

The XDP write path aggregates random writes in a small (64 KB) host arrival buffer that is periodically flushed to a capacitor-backed onboard buffer. Upon reaching capacity, the latter is indexed, compressed, and flushed to SSD storage in big (1--2 GB) stripes. Every value is stored together with its hash key. Within each stripe, the data is ordered by database id and hash key, so the values within the same database are clustered per-stripe but not ordered by user key. Clustering by database id serves whole-database scans, which use sequential I/O. Intra-cluster ordering by hash key serves memory-efficient hashtable encoding.

The hashtable is compressed -- it reduces 32-bit hash keys to small variable-length fingerprints. Its bits-per-key ratio is close to the minimum required for unique decoding. But compression comes at a cost -- updating a non-empty slot in the table requires testing if the colliding fingerprint maps to an existing key or a new one. Therefore,  when flushing a stripe, XDP re-reads every colliding fingerprint to discover the full key. If that key is different from the new key, the whole slot is re-encoded. This background read is called \emph{fetch-existing-entry}, or \emph{fee}. To reduce the performance cost of fees,  XDP's put and delete APIs provide an overwrite hint to indicate whether the update applies to an existing key.

XDP endures  host surprise power-down failures through a shutdown procedure that drains the buffered data to storage. It also guarantees data integrity in case of application crash through a recovery mechanism. The maximal data loss is limited to the host arrival buffer size.

\subsection{XDP-Rocks architecture}
\label{sec:xdprocks}

This section focuses on how XDP-Rocks addresses real-life challenges and compares its implementation details to RocksDB's. For efficient storage management, the LSM  is stored in XDP via a custom KVS filesystem (KVFS), as described in Section \ref{sec:kvfs}. Scans are accelerated using parallel random-read workers, as explained in Section \ref{sec:xdprocks-scan}. We discuss the XDP-Rocks cache structure in Section \ref{sec:caching}, and conclude with checkpoints and backup in Section \ref{sec:checkpoint}. For lack of space we skip other implementation details, e.g., supporting the RocksDB's compaction filter API~\cite{CompactionFilter} and tuning its rate limiter~\cite{RateLimiter} for KV-separation.

\subsubsection{KVFS -- the key-value filesystem}
\label{sec:kvfs}

A significant system management challenge in KV-separated LSMs 
is allocating capacity to the LSM and the KVS over the same physical device(s). A straightforward approach could pre-partition the storage space between the LSM (keys) and the values in advance. But doing so without wasting storage space requires advance knowledge about the key-to-value size ratio statistics. Moreover, LSM space allocation  must accommodate for temporary SA during compactions~\cite{DayanWDPBT22}. 
XDP-Rocks instead uses XDP to store both the LSM and the values, without pre-partitioning.

To facilitate data storage over XDP, XDP-Rocks implements \emph{KVFS} -- a KVS filesystem. 
Every XDP-Rocks database maps to two distinct XDP databases -- one for values and one  for  LSM (SST and WAL) files. The latter 
uses RocksDB's virtual-filesystem mechanism~\cite{DongPPAESDPJKPZ23}, which maps the LSM directory to the XDP database.
File sizes are flexible, and every file holds  metadata that links it to its XDP database.
A database drop translates to deleting the two XDP databases and the top-level directory.

KVFS is tailored to the LSM write-once-read-many I/O pattern. 
XDP-Rocks writes SST and WAL files to KVFS as sequences of approximately equal-sized logical blocks (4 KB for SST and 32 KB for WAL). Every read retrieves a sequence of one or more  blocks. KVFS handles this workload by storing every file as a set of KV-pairs, each standing for a single block. The KV-pairs are organized in variable-size \emph{extents}; a KVFS file is a single extent. The keys used by KVFS to store blocks in the same extent share a common prefix (extent id), and their suffix is a running index. 

Allocating new extent ids for new files comes with a cost -- it causes writes to use new keys 
for KVFS blocks, which triggers a background read-after-write in XDP (fee, see Section~\ref{sec:xdp}). To forgo this cost, KVFS recycles  extent keys. Every time a file is deleted, its extent id returns to the free pool, and whenever a new file is created, KVFS first tries to reuse an existing extent id. For the reused keys, KVFS applies an overwrite hint, which avoids the fee overhead. This boosts the KVFS write performance by approximately 20\% versus the na\"ive design. 
An extra challenge is ACID maintenance of the extent pool metadata, to address crash recovery. This is achieved through the RocksDB manifest (metadata log) mechanism~\cite{RocksDB}. 

\subsubsection{Scan performance}
\label{sec:xdprocks-scan}

Range reads (scans) pose the biggest challenge for XDP-Rocks due to the use of an unordered KVS. Although our target workload uses them sparingly, we still need to offer reasonable performance
for those that do occur. Our solution is similar to WiscKey's~\cite{wiscKey} -- expedite scans through parallel reads. Namely, XDP-Rocks employs a background worker thread pool to read multiple values in parallel, coupled with a concurrent queue to handle out-of-order completions. The maximum number of workers is configurable per-iterator.  

In addition to value-read parallelism, XDP-Rocks uses the same mechanism to facilitate readahead in KVFS sequential reads. Namely, multiple threads (4 by default) prefetch sequential KVFS blocks in parallel. This pipelining reduces the number of stalls in application-level scans as well as in compactions and recovery. 
Note that these  mechanisms  cannot fully match the sequential read performance of baseline RocksDB, which relies on standard filesystem readahead (2 GB, by default), especially as the scans scale (Section~\ref{sec:microbench}).


\subsubsection{Data caches}
\label{sec:caching}

Recall that XDP does not have a cache of its own (Section~\ref{sec:xdp}), so it is up to XDP-Rocks to cache hot values. 
Like RocksDB, XDP-Rocks uses two types of caches. The first is a \emph{block cache} holding LSM SST file blocks to speed up LSM searches. The second is a \emph{row cache} that holds (variable-size) values rather than (fixed-size) SST data blocks. The latter supports random access with better memory efficiency, unless values are small, in which case the per-value overhead is dominant. The Bloom filters and the SST index blocks of all valid LSM files are permanently pinned in memory as part of the block cache in both systems.


XDP-Rocks's row cache is more efficient than RocksDB's because RocksDB does not have direct storage of values by key, so its row cache holds popular values per SST file (the file id is part of the key). When a cached value becomes obsolete in case of update or deletion, its eviction from the row cache is lazy, to avoid explicit search of the affected file. In contrast, XDP-Rocks caches values under their user keys and updates them in-place. 

\subsubsection{Checkpoints and backup}
\label{sec:checkpoint}

The RocksDB \code{checkpoint(dir)} API in XDP-Rocks creates a persistent point-in-time view of an entire database in the  directory \code{dir}. The checkpoint can then be used, e.g., by backup tools, to replicate the database \cite{BackupRocksDB}. In XDP-Rocks, we restrict ourselves to read-only checkpoints, which are the most common use case; the primary replica remains read-write.

An XDP-Rocks checkpoint is a database that stores a copy-on-write (CoW) clone of the primary replica's LSM tree under \code{dir} and shares its KVS storage (values and KVFS) with the primary. XDP-Rocks leverages the snapshot API to capture a virtual copy of the primary's storage at the time of checkpoint creation. Once a snapshot is created, updates to the primary do not overwrite values that pertain to the checkpoint. XDP-Rocks persists the snapshot's \code{sn} timestamp, and re-installs that snapshot every time the primary database is reopened. The checkpoint's deletion de-persists its \code{sn} and removes the snapshot from the primary. Note that RocksDB's checkpoints are simpler than XDP-Rocks's: they only use the CoW mechanism for LSM 
and do not require snapshots for retaining values (because RocksDB is not KV-separated).

Long-lived checkpoints degrade performance because they cause all subsequent writes to the primary database to be versioned, which effectively disables the LSM bypass. They also waste space for keeping snapshot-protected values, albeit much less than RocksDB, which doubles storage through CoW as the primary LSM evolves.

In principle, XDP-Rocks's backup could have been created via the standard RocksDB API, by scanning the whole checkpoint database and re-inserting its KV pairs into the initially empty target database one-by-one.
However, this approach would have been inefficient, due to inherent scan performance challenges in KV-Tandem. Speeding up scans through internal parallel reads (Section~\ref{sec:xdprocks-scan}) offers a partial remedy but is ultimately inefficient for copying terabytes of data. To support efficient backup, XDP-Rocks relies on file-level copy of the checkpoint LSM (as RocksDB does~\cite{BackupRocksDB}), coupled with efficient copy of the underlying KV-separated values. 

Since checkpoints are read-only, the data can be copied in any order -- for instance, the order of values needs not match the order of their keys. The copy process streams the two datasets into two initially empty XDP databases (values and KVFS) at the target and complements them with a copy of \code{dir}, the  checkpoint directory atop KVFS. At the end of this bottom-up construction, the target holds a consistent XDP-Rocks database. The LSM-copy uses KVFS whole-file scan, which is slower than RocksDB's (Section~\ref{sec:xdprocks-scan}) but still reasonably fast because the LSM is small. The value-copy scans the KVS value database (which is shared with the primary) via XDP's whole-database out-of-order scan API that that exploits sequential I/O to its full extent (Section~\ref{sec:xdp}).

At the end of the copy process, the backup storage might hold newer writes that occurred to the primary after the snapshot. To trim the backup, we scan the primary's WAL starting from the snapshot \code{sn} and delete the new writes from the backup's KVS, via the standard API.


\section{Evaluation}
\label{sec:results}
We evaluate XDP-Rocks  extensively in settings that reflect a multi-terabyte production deployment (to the best of our knowledge, this is the largest-scale study of this kind). Section~\ref{ssec:methodology} describes our methodology and experiment environment. 
In Section \ref{ssec:eval-kv-separation}, we compare XDP-Rocks to BlobDB~\cite{BlobDB}, a RocksDB variant that implements the WiscKey KV-separation method. We highlight BlobDB's unbounded SA and subsequently omit it from further experiments. Section~\ref{sec:microbench} presents  storage engine microbenchmarks and Section~\ref{ssec:eval-db} presents a system-level evaluation of the Kvrocks NoSQL database~\cite{KVRocks}, which runs atop RocksDB-compatible storage.

\subsection{Methodology}
\label{ssec:methodology}

\paragraph{Compared systems}
We compare XDP-Rocks to RocksDB 9.3 (released May 25, 2024).
As a performance upper bound for random reads and writes, we also compare our solution -- where applicable -- to XDP, the underlying unordered KVS.  In addition, to understand the algorithmic contribution of KV-Tandem over the basic KV-separation concept,  we implement a KV-separated LSM, called \emph{XDP-Rocks-Nodirect}, which uses XDP to store values in versioned mode only (with no direct storage and no LSM bypass). 

\paragraph{Hardware and OS} 
We use a Dell PowerEdge R750 server with a 64-core Intel Xeon Gold 6346 CPU, 503 GB DRAM, XDP Flex1 FPGA and four Samsung PM9A3 SSDs (total capacity 15.36 TB). 90\% of the SSD capacity is allocated to XDP (approximately, 13.7 TB). The server is deployed with Linux kernel 4.15 (Ubuntu 18.04.6). 

\paragraph{Data} Unless specified otherwise, the experiment creates and runs its workloads over 16 storage-engine databases -- i.e., a system that manages 16 data shards. 
We fill the storage with 9.2 TB of data (approximately $70\%$ of the storage) with sequentially-growing keys. This is the most RocksDB can scale to until running out of space. The size of every key is 32 B and the size of every value is 1 KB; the values are non-compressible. 

\paragraph{Workloads} 
For our microbenchmarks, we use  \emph{dbbench}~\cite{RocksDB}, slightly modified to run  workloads resembling the standard ones in the YCSB suite~\cite{CooperSTRS10}. We run four workloads: write-only, read-only (aka YCSB-C), mixed (50\% read, 50\% write -- aka YCSB-A), range query (scan), and scan-write (aka YCSB-E). All writes are updates of existing keys.
We also experimented with a read-heavy workload (95\% read, 5\% write -- aka YCSB-B); the results were similar to those of the read-only benchmark and so are omitted. We run all workloads with a uniform key distribution. In cases of interest, we also run a Zipfian workload with power parameter  $\alpha=1.2$ (it is not applicable to XDP, which by design has no cache.)

\paragraph{Experiment setup and predictability} 
We generate every workload through a varying number of worker threads (as long as the system keeps scaling). 
Every operation is routed to one of the 16 databases at random. Every experiment (for a given workload and number of threads) begins with a full database and runs for one hour. 
To avoid  cross-workload noise (effects of past experiments on the measured results), we use guard periods. To avoid post-fill effects (e.g., compaction rate above the stable-state due to an unbalanced LSM~\cite{GiladBBGHKMS20}), we run uniform updates of existing keys through 16 threads for 4 hours after the fill completes. 

\paragraph{Software configuration} 
To crystallize the effect of I/O, we run all systems without a block cache, except the pinned Bloom filter and index blocks.  
In uniform workloads, we do not employ any cache at all, whereas in Zipfian ones, we use a row cache of 300 GB.  
Recall that in uniform workloads, the cache has little benefit, and in Zipfian workloads with large values, the highest I/O reduction is achieved by the row cache.   

We use 16 background worker threads for flushes and 16 more for compactions, to allow independent progress across all databases. The LSM branching factor is 10 and the LSM compaction policy is RocksDB's leveled compaction~\cite{RocksDB}. Both RocksDB and XDP-Rocks employ RockdsDB's dynamic capacity allocation (DCA) policy to control the LSM SA~\cite{DongCGBSS17}. We use the asynchronous WAL option. The rest of the system parameters use default values. 

\subsection{Limitations of KV-separation solutions}
\label{ssec:eval-kv-separation}

\begin{figure}[htb]
    \centering
      \includegraphics[width=0.35\textwidth]{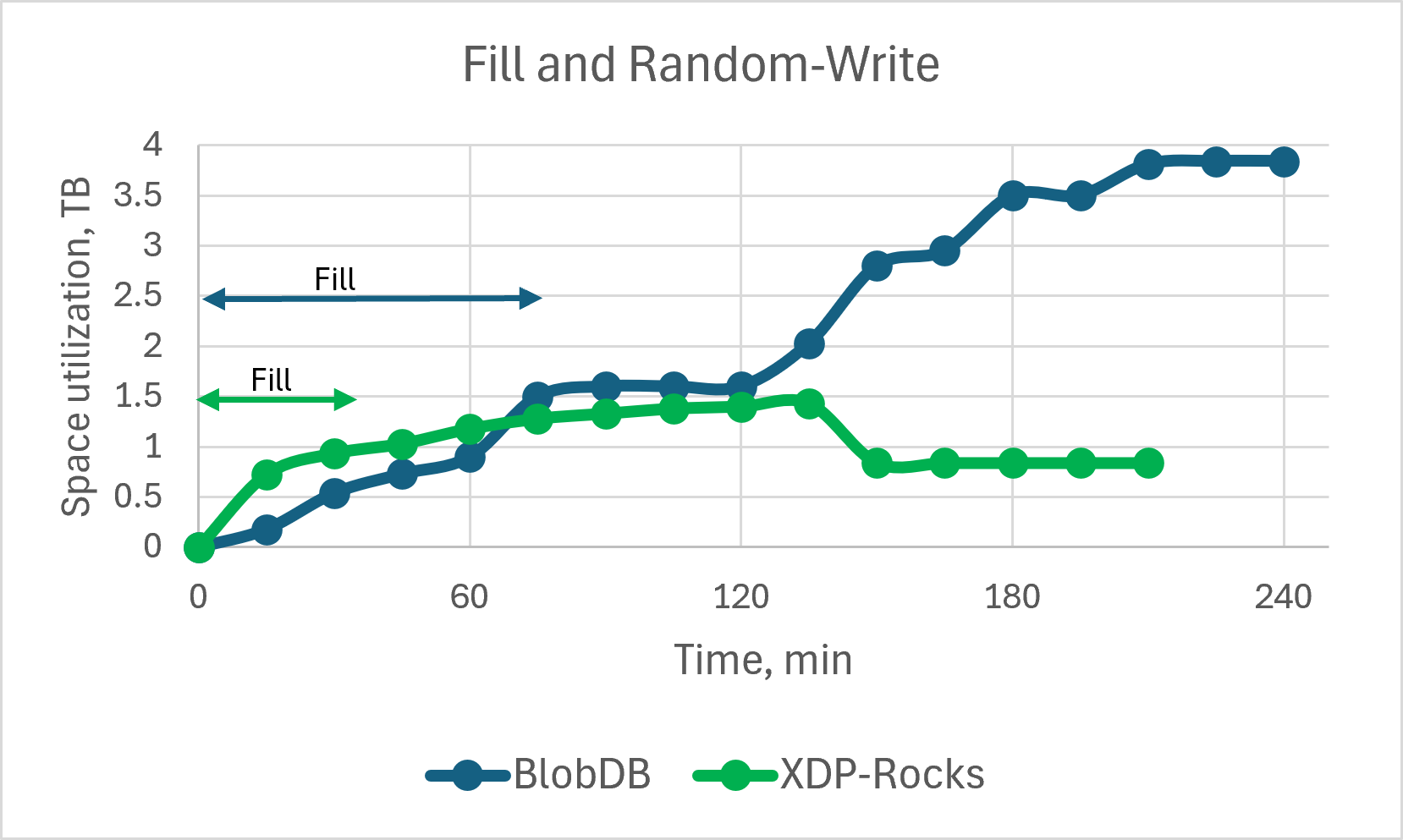}
       \caption{BlobDB's unbounded storage growth.}  
      \label{fig:blobdb-capacity}
\end{figure}

We start with a small experiment that highlights the severe SA limitation of  BlobDB, 
a RocksDB variant following WiscKey's KV-separation design. Due to tight coupling between LSM compaction and value-log GC, BlobDB cannot sustain a large dataset. Our setup is as described in Section~\ref{ssec:methodology}, but at a smaller scale: We use a single SSD of capacity 3.84 TB. We only fill the storage with 650 GB of data and then run a 3-hour random-write period. Figure~\ref{fig:blobdb-capacity} depicts the storage space utilization over time, starting from an empty drive.

XDP-Rocks and BlobDB complete the fill stage within 35 and 75 minutes, respectively. Both  utilize approximately 1 TB of the SSD space at that point. At the beginning of the random write period, the XDP-Rocks space utilization  keeps growing and peaks at 1.4 TB. At that point, the XDP GC wakes up and collects the invalidated space, thus reducing the use to 840 GB. In contrast, BlobDB's storage footprint grows without bound until running out of storage space completely.

\subsection{Microbenchmarks} 
\label{sec:microbench}

We now use microbenchmarks to evaluate XDP-Rocks versus RocksDB in a variety of intensive workloads. 
We compare the storage engines in terms of throughput (queries per second, or qps), tail ($99.99\%$) latency under a cutoff of 10 ms, and space utilization. 

\subsubsection{Random write}

\begin{figure}[h]
  \centering
    \begin{minipage}{0.4\textwidth}
        \centering
        \includegraphics[width=\textwidth]{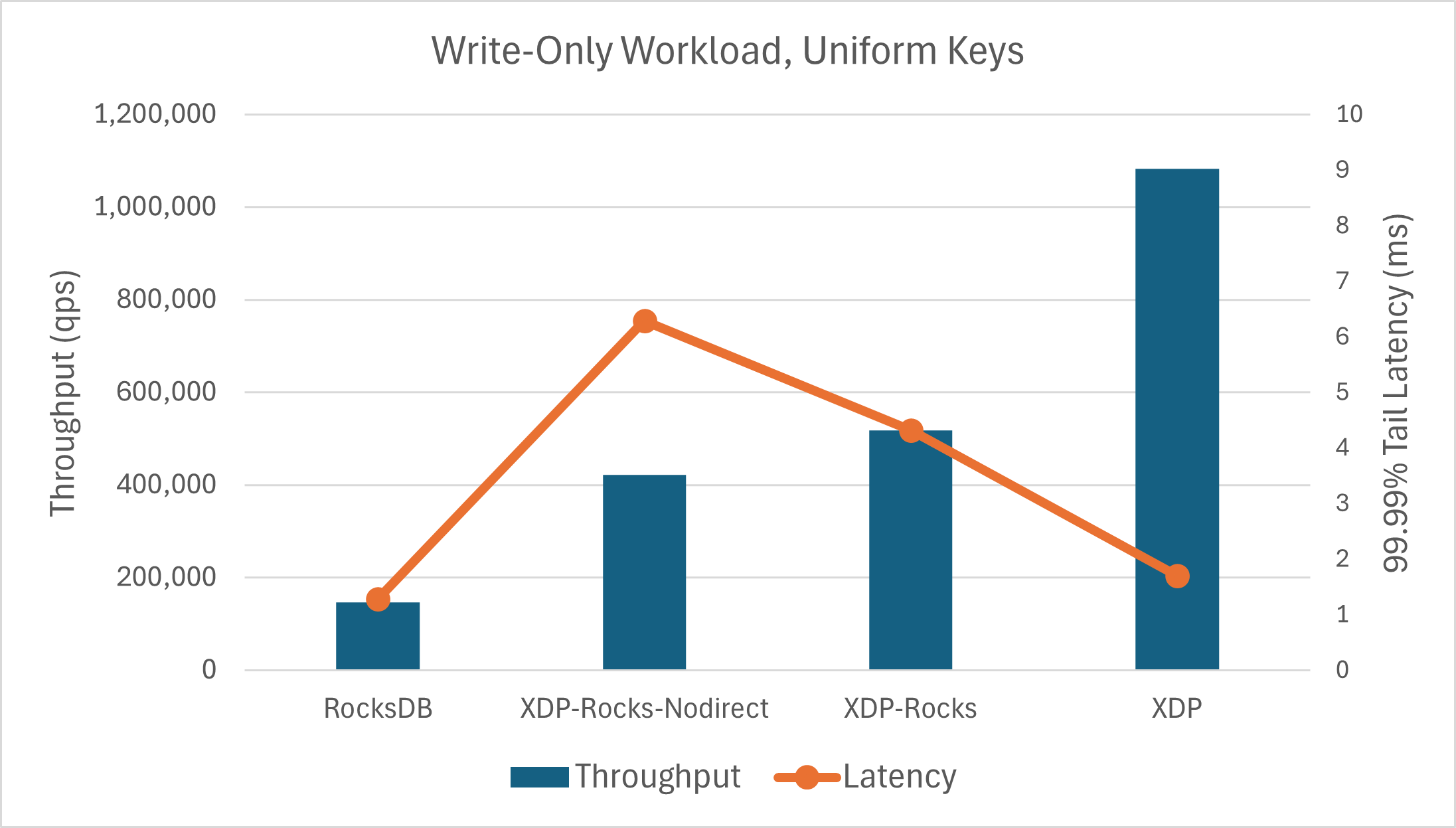}
        \subcaption{Maximal throughput and latency}
        \label{fig:ycsb-o-throughput}
    \end{minipage}
    \hfill
    \begin{minipage}{0.4\textwidth}
        \centering
        \includegraphics[width=\textwidth]{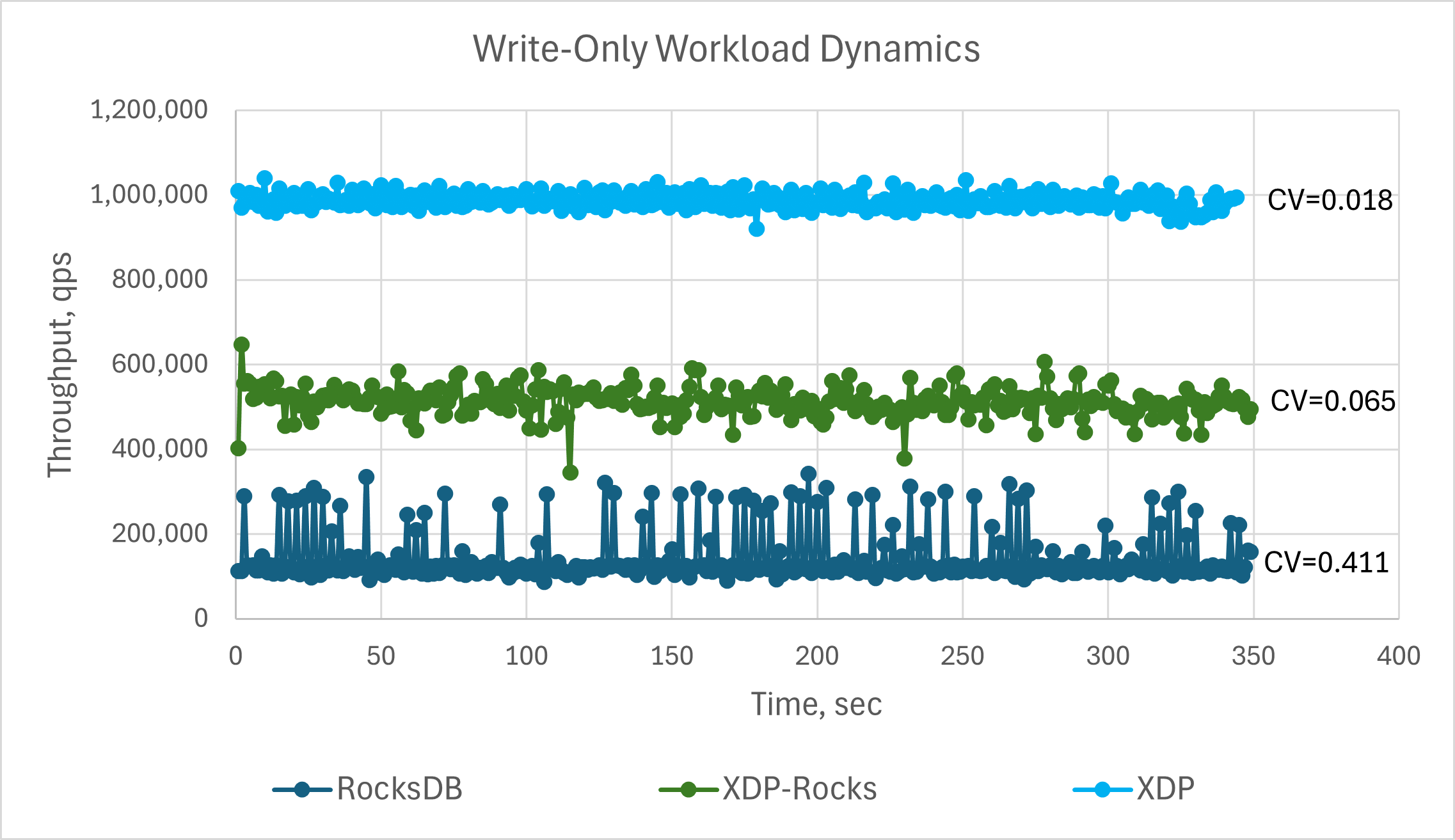}
        \subcaption{Throughput dynamics}
        \label{fig:ycsb-o-ts}
    \end{minipage}
\caption{Random write-only (update) workload.}
\label{fig:ycsb-o-workload}
\end{figure}

The results of  the update-only workload are given in Figure \ref{fig:ycsb-o-workload}. 
We see that XDP-Rocks achieves approximately 520K qps -- 3.5x above RocksDB and 1.23x above XDP-Rocks-nodirect (Figure~\ref{fig:ycsb-o-throughput}). The advantage of XDP-Rocks-nodirect over RocksDB stems from KV-separation,
which substantially reduces the WA and thus improves the I/O bandwidth utilization.  

Interestingly, KV-Tandem, which had  originally been designed to improve  read performance, also improves write performance.
This stems from reducing the number of calls to the API of the underlying KVS. In XDP-Rocks-nodirect, all KVS entries are versioned, and so every write leads to two KVS calls -- one to create a new version and one to remove the previous one (the latter occurs during compaction). While the delete calls do not normally induce extra I/O, they are associated with a compute overhead, manifesting in slightly reduced performance. 

The performance gap between XDP-Rocks and the baseline XDP is roughly 2.09x. This stems primarily from writing to WAL -- every write to XDP-Rocks is first written to WAL (from memtable) and only then added to the storage. This accounts for 2x WA. The extra 9\% is due to the LSM's WA for keys. The maximum throughput was achieved with 2--4 threads in all systems, due to RocksDB's non-scalable memtable implementation~\cite{Golan-GuetaBHK15}.

\remove{
In contrast with the ordered storage engines, XDP achieves 1.09M qps -- approximately 2.09x versus XDP-Rocks. This measurement closely follows the XDP-Rocks's WA model (see Section~\ref{sec:implementation} for analysis) and serves as upper bound for theoretically achievable performance. The raw I/O numbers corroborate the analysis directly. XDP observes a write rate of 1.17 GB/s under XDP-Rocks, which is approximately 2.25x of the application's data rate. 

The total write bandwidth projected towards the storage is 1.97 GB/s (i.e., the GC rate is 0.8 GB/s). This is still far from the nominal SSD sequential write bandwidth (4 GB/s~\cite{PM9A3}, or 16 GB/s system-wide). We attribute this to (1)  XDP Flex1 scalability gaps and (2) the \emph{fee} reads  triggered by this workload (Section~\ref{sec:implementation}). Namely, every direct-value write triggers a background fee read, i.e., the SSDs handle 520K random read iops in parallel with the sequential writes. 
}

Figure~\ref{fig:ycsb-o-ts} zooms in on the performance dynamics. The number next to each curve is the \emph{coefficient of variation} (\emph{CV}), namely, the STD-to-mean ratio. 
RocksDB is extremely spiky with a CV of 41.1\%, as its compactions constantly throttle the system (this behavior is well-known~\cite{LuoC19a}).
At the other extreme, the baseline XDP is very predictable (1.8\% CV).
XDP-Rocks is fairly stable (6.5\% CV), though less than XDP. The fluctuations stem from the same compactions as in RocksDB, albeit less significant, as the LSM is much smaller. We omit XDP-Rocks-nodirect to avoid obscuring the graph; its dynamics are similar to those of XDP-Rocks.

Finally, we note that XDP-Rocks has near-zero space overhead (SA close to 1): it utilizes 9.3 TB of storage for 9.1 TB data. In contrast, both XDP-Rocks-Nodirect and RocksDB utilize 10.2 TB of storage -- nearly 10\% space overhead. This is explained as follows. XDP-Rocks overwrites direct values in XDP upon flush, and so XDP's GC can collect  stale overwritten values promptly. On the other hand, XDP-Rocks-Nodirect  deletes stale versioned values only at compaction time. Therefore, its stable-state SA is the same as the LSM's (and consequently, RocksDB's too). It is known that with an LSM branching factor of $f$, the expected space amplification under leveled compaction is $1/(f-1)$~\cite{DayanWDPBT22} -- a tight match for $f=10$.

\subsubsection{Random read}

\begin{figure}[h!]
  \centering
    \begin{minipage}{0.4\textwidth}
        \centering
        \includegraphics[width=\textwidth]{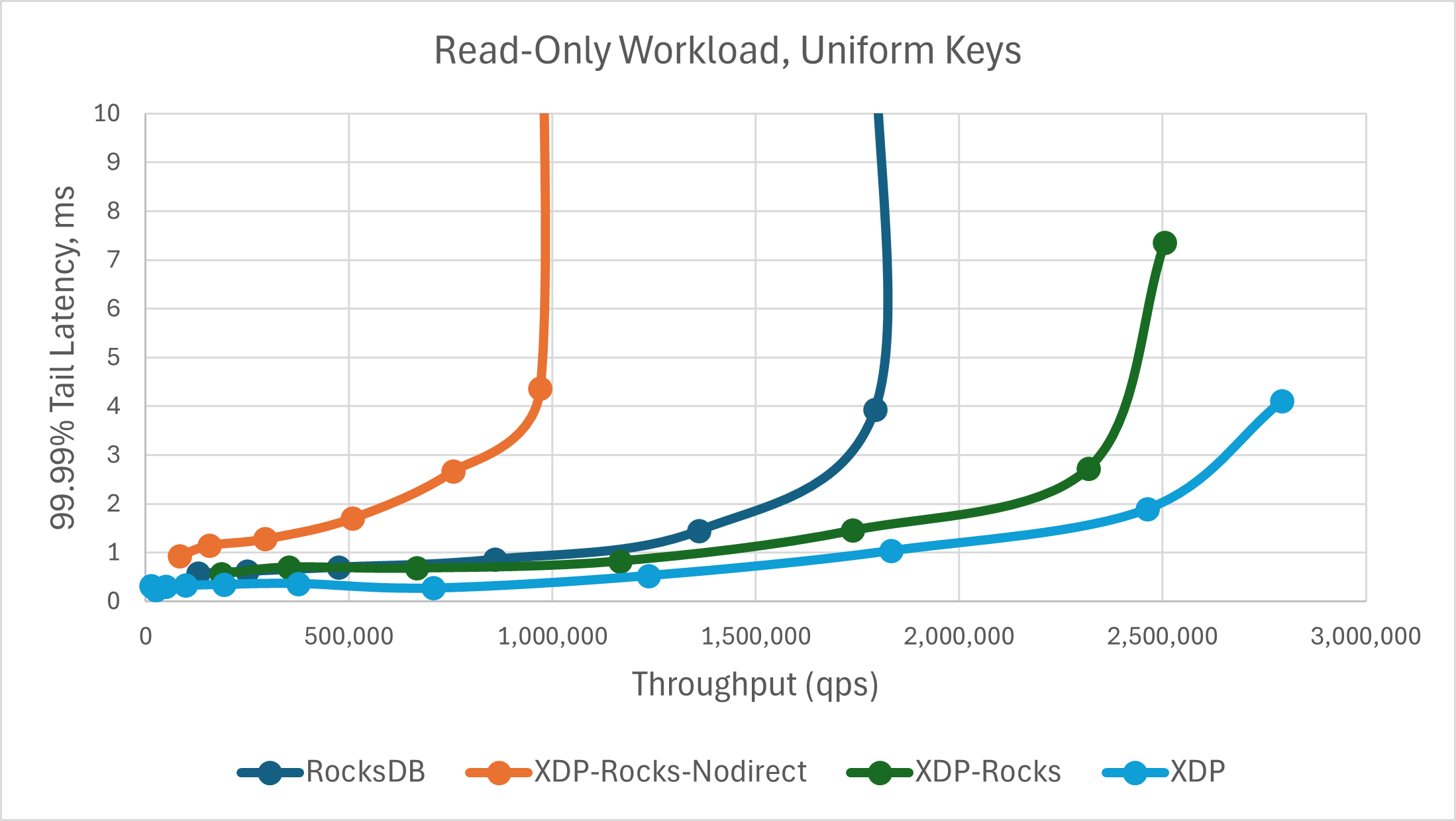}
        \subcaption{Uniform distribution}
        \label{fig:ycsb-c-u}
    \end{minipage}
    \hfill
    \begin{minipage}{0.4\textwidth}
        \centering
        \includegraphics[width=\textwidth]{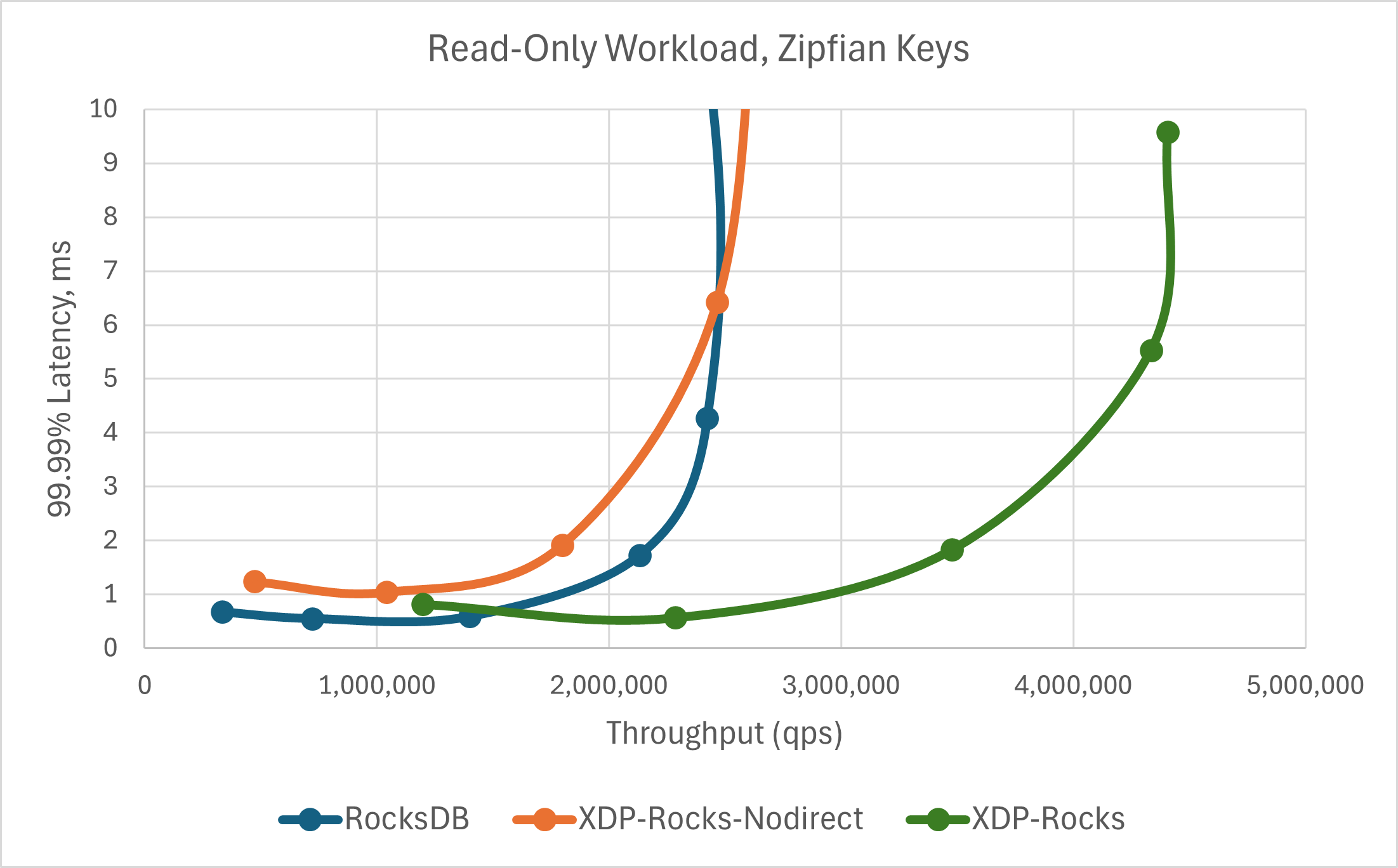}
        \subcaption{Zipfian distribution}
        \label{fig:ycsb-c-z}
    \end{minipage}
\caption{{Random read-only workload.}}
\label{fig:ycsb-c-workload}
\end{figure}

Figure \ref{fig:ycsb-c-workload} shows throughput versus latency results of the 100\%-read workload scaling (aka YCSB-C) to 1024 threads.  
First, we look at a uniform key distribution (Figure~\ref{fig:ycsb-c-u}). Here, XDP-Rocks scales to 2.5M qps, very close to the upper bound of 2.79M qps achieved by XDP. Note that XDP-Rocks calls XDP's get exactly once per read, so the two systems perform the same amount of I/O and the minor gap between them is due to compute overhead. Each XDP get reads a 1 KB value. Because values in XDP are not aligned, the get leads to reading one 4 KB physical block with probability 75\% and two blocks with probability 25\%, or 1.25 blocks in expectation. 
The I/O in RocksDB is structured differently. Each SST block is 4KB-aligned, i.e., covers two physical blocks (expected 8KB SSD read). Thus, the read throughput can be expected to be roughly 62.5\% that of XDP-Rocks. Indeed, RocksDB scales to 1.79M qps, which is approximately 64\% of the XDP throughput. 
XDP-Rocks-Nodirect scales to 970K qps because it reads both the SST block and the value ($1.25+2=3.25$ blocks), hence its RA versus XDP-Rocks is 2.6x, matching the measured gap precisely. 

Figure~\ref{fig:ycsb-c-z} compares the systems under a Zipfian workload. XDP-Rocks and XDP-Rocks-Nodirect scale to 4.4M qps and 2.5M qps, respectively. RocksDB is slightly faster than the latter. All the systems enjoy the row cache optimization, with an identical steady-state hit rate (75\%); caching reduces the I/O-driven performance gaps.

\subsubsection{Mixed write and read}

\begin{figure}[ht]
  \centering
    \begin{minipage}{0.4\textwidth}
        \centering
        \includegraphics[width=\textwidth]{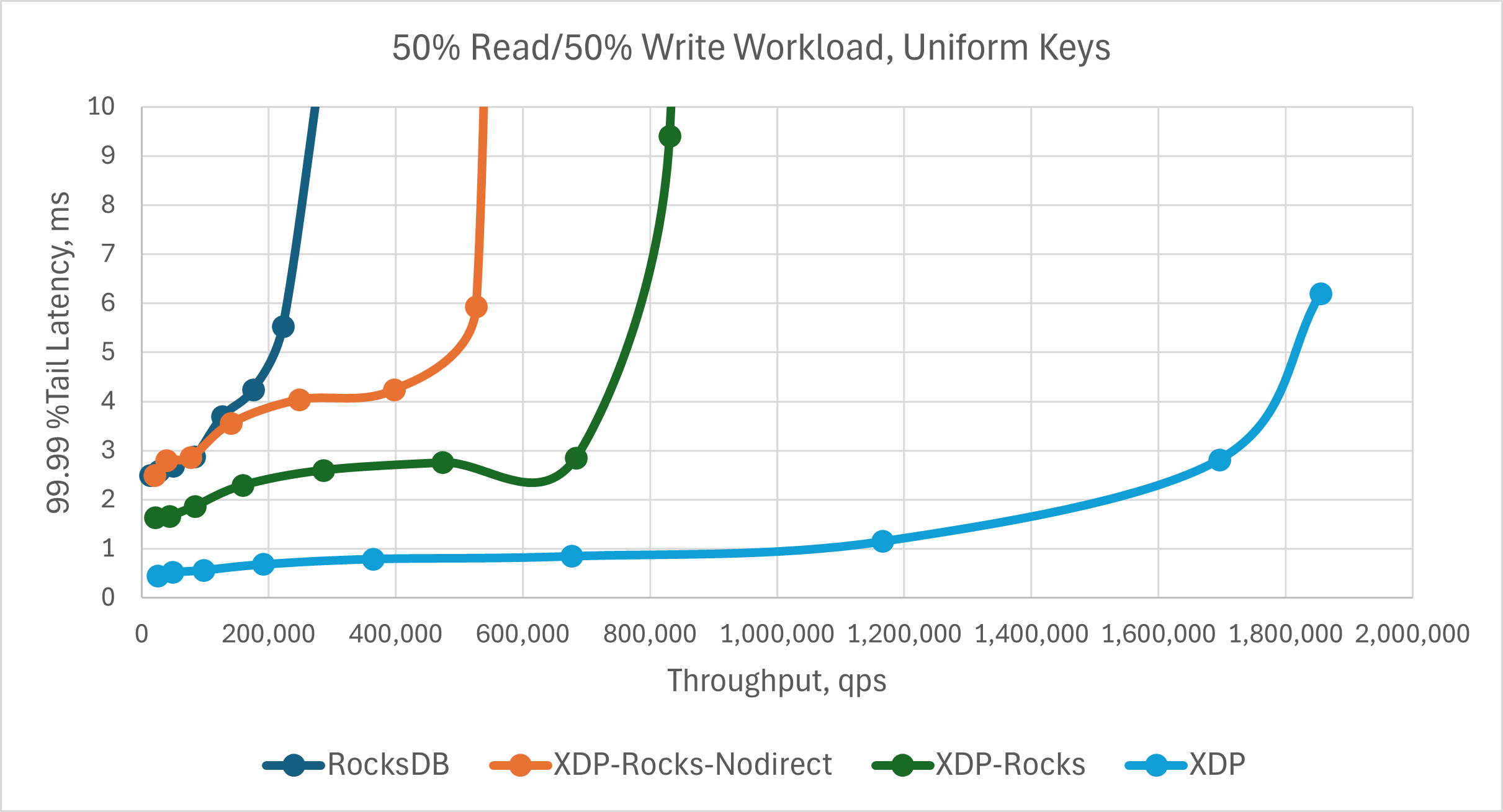}
        \subcaption{Uniform distribution}
        \label{fig:ycsb-a-u}
    \end{minipage}
    \hfill
    \begin{minipage}{0.4\textwidth}
        \centering
        \includegraphics[width=\textwidth]{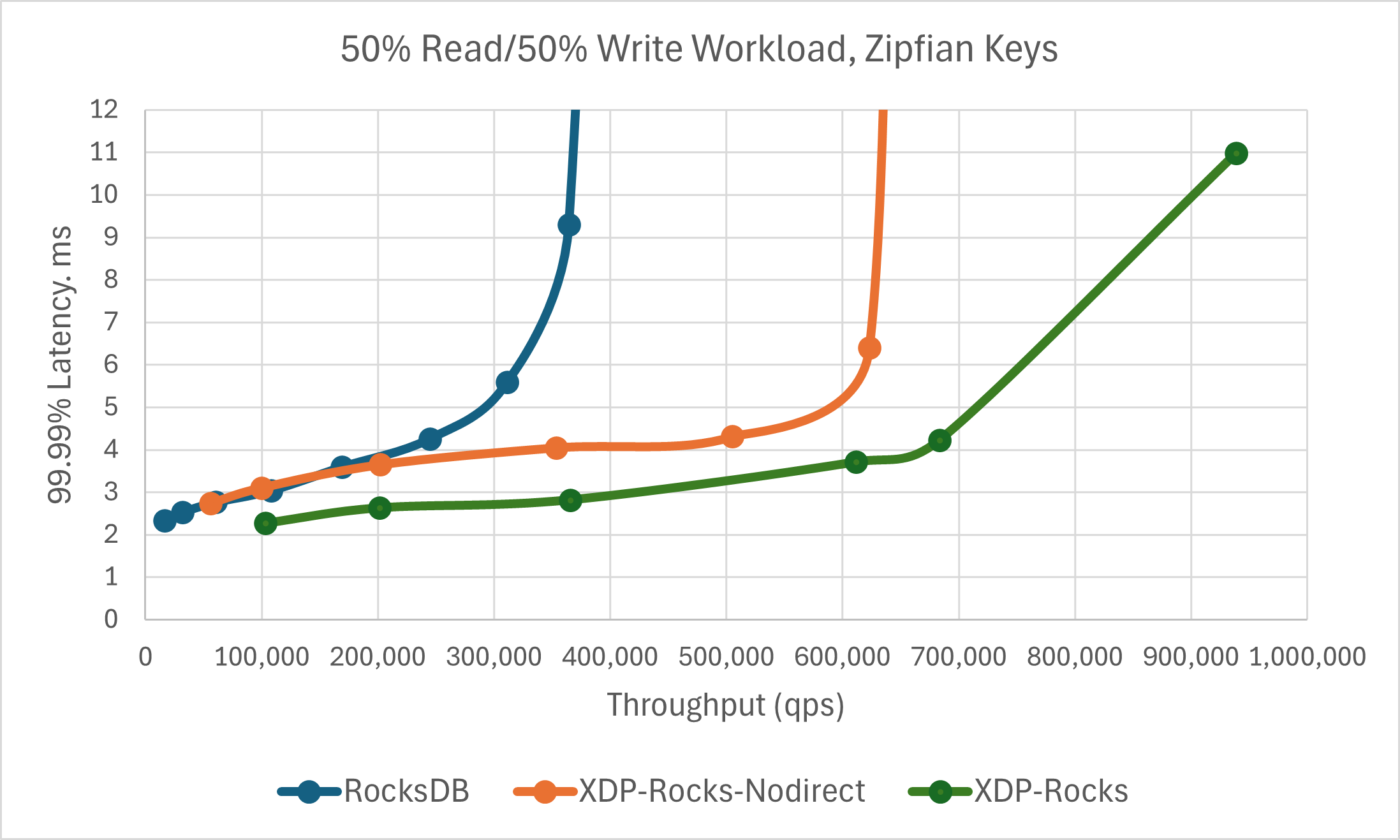}
        \subcaption{Zipfian distribution}
        \label{fig:ycsb-a-z}
    \end{minipage}
\caption{{Mixed workload: 50\% read, 50\% write.}}
\label{fig:ycsb-a-workload}
\end{figure}

Figure \ref{fig:ycsb-a-workload} studies the write-heavy mixed (50\% write/50\% read) workload (aka YCSB-A). The performance under the uniform distribution (Figure~\ref{fig:ycsb-a-u}) aligns closely with the 100\%-write results, as the write path dominates all the systems. Here, XDP-Rocks exceeds RocksDB by 3.8x (940K qps versus 430K), whereas XDP scales to 1.86M qps. Under the Zipfian distribution (Figure~\ref{fig:ycsb-a-z}), the gap between XDP-Rocks and RocksDB reduces to 2.2x since a big fraction of reads is now served from the row cache. Note, however, that this gap is bigger than under the read-only workload (1.8x), thanks to XDP-Rocks's better cache utilization and hit rate. Recall that under the mixed workload the RocksDB cache may store multiple versions per key and suffer from unnecessary invalidations, whereas the XDP-Rocks cache stores at most one version per key and gets updated proactively (see Section~\ref{sec:implementation}).

\subsubsection{Scan and scan-write}

\begin{figure}[htb]
        \centering
      \includegraphics[width=0.35\textwidth]{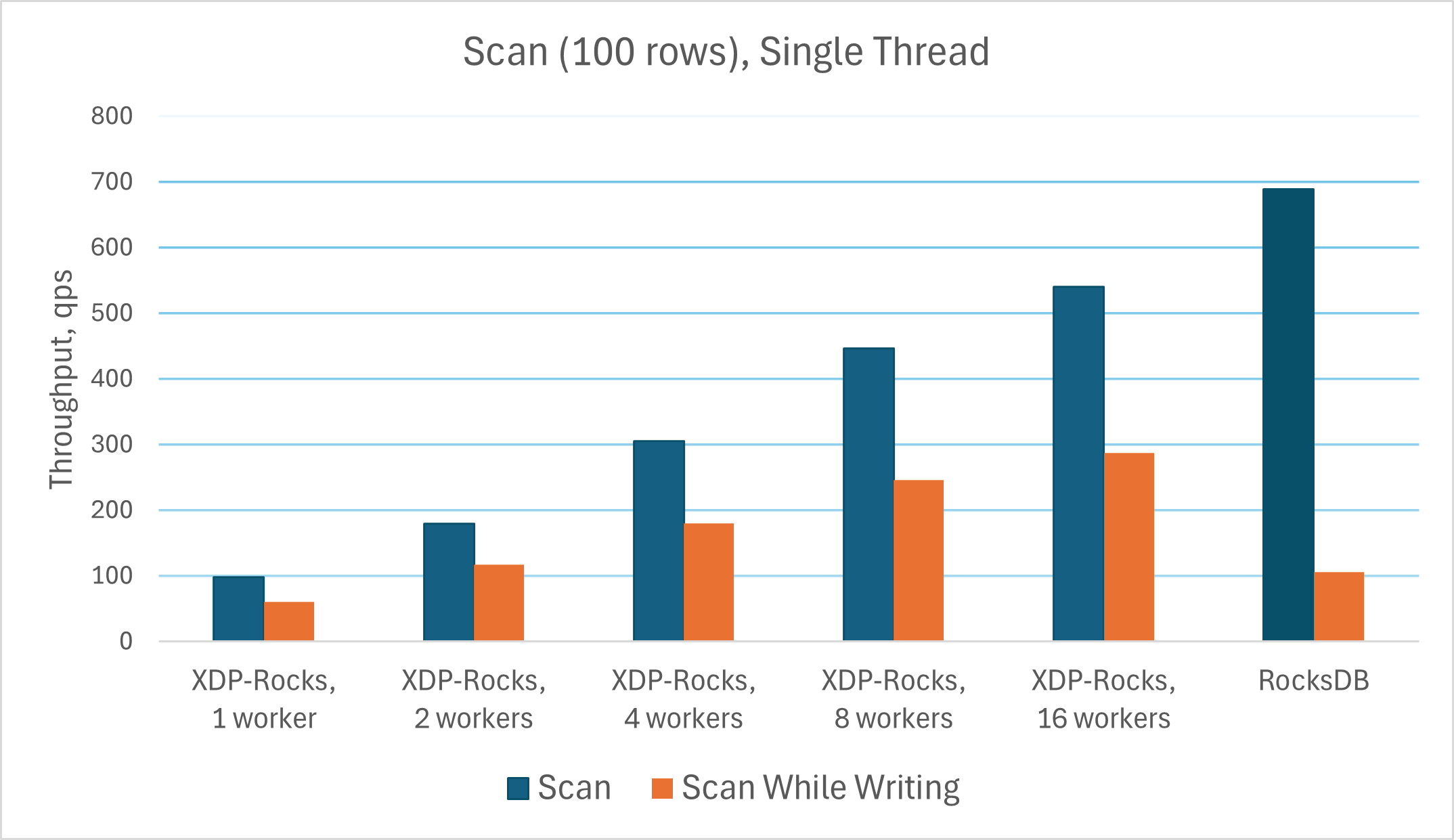}
       \caption{Single scan with and without background writes.}  
      \label{fig:scan-single-thread}
\end{figure}

In this section we measure scan performance, both without background updates (the scan workload) and with them (the scan-write workload).
As described in Section \ref{sec:implementation}, we expedite scans by using multiple worker threads to perform the scan in parallel. Figure~\ref{fig:scan-single-thread} depicts the performance of a single scanning application thread supported by 1  to 16 background worker threads. The application thread repeatedly selects a key uniformly at random  and then iterates through the next 100  keys. In the scan experiment (blue bars), the scanning thread is the only application thread, whereas in the scan-write experiment (orange bars), one more application thread concurrently executes writes. 

\begin{figure}[htb]
    \begin{minipage}{0.4\textwidth}
        \centering
        \includegraphics[width=\textwidth]{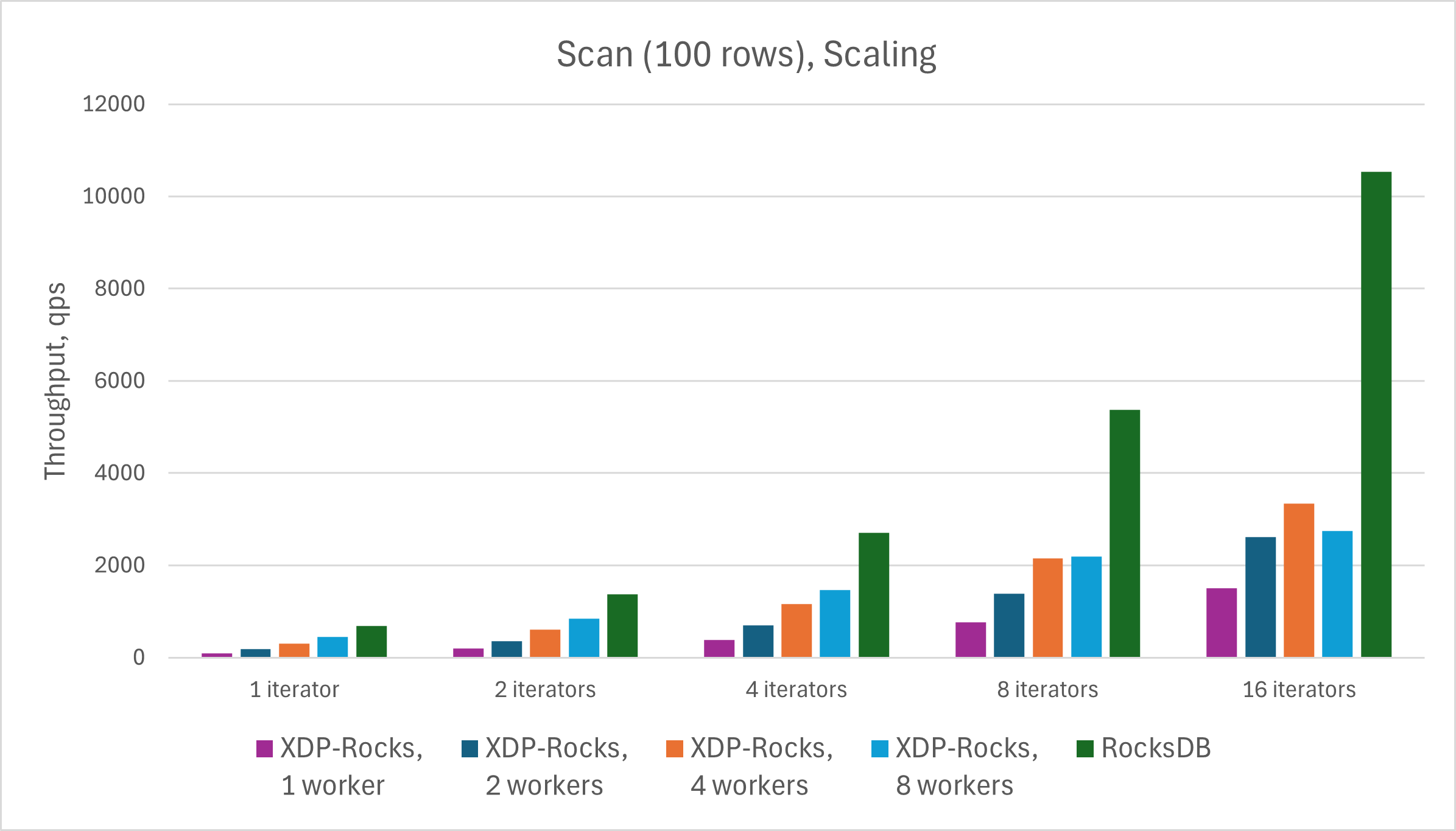}
        \subcaption{Scan scaling}
        \label{fig:scan-scaling}
    \end{minipage}
    \hfill
    \begin{minipage}{0.4\textwidth}
        \centering
        \includegraphics[width=\textwidth]{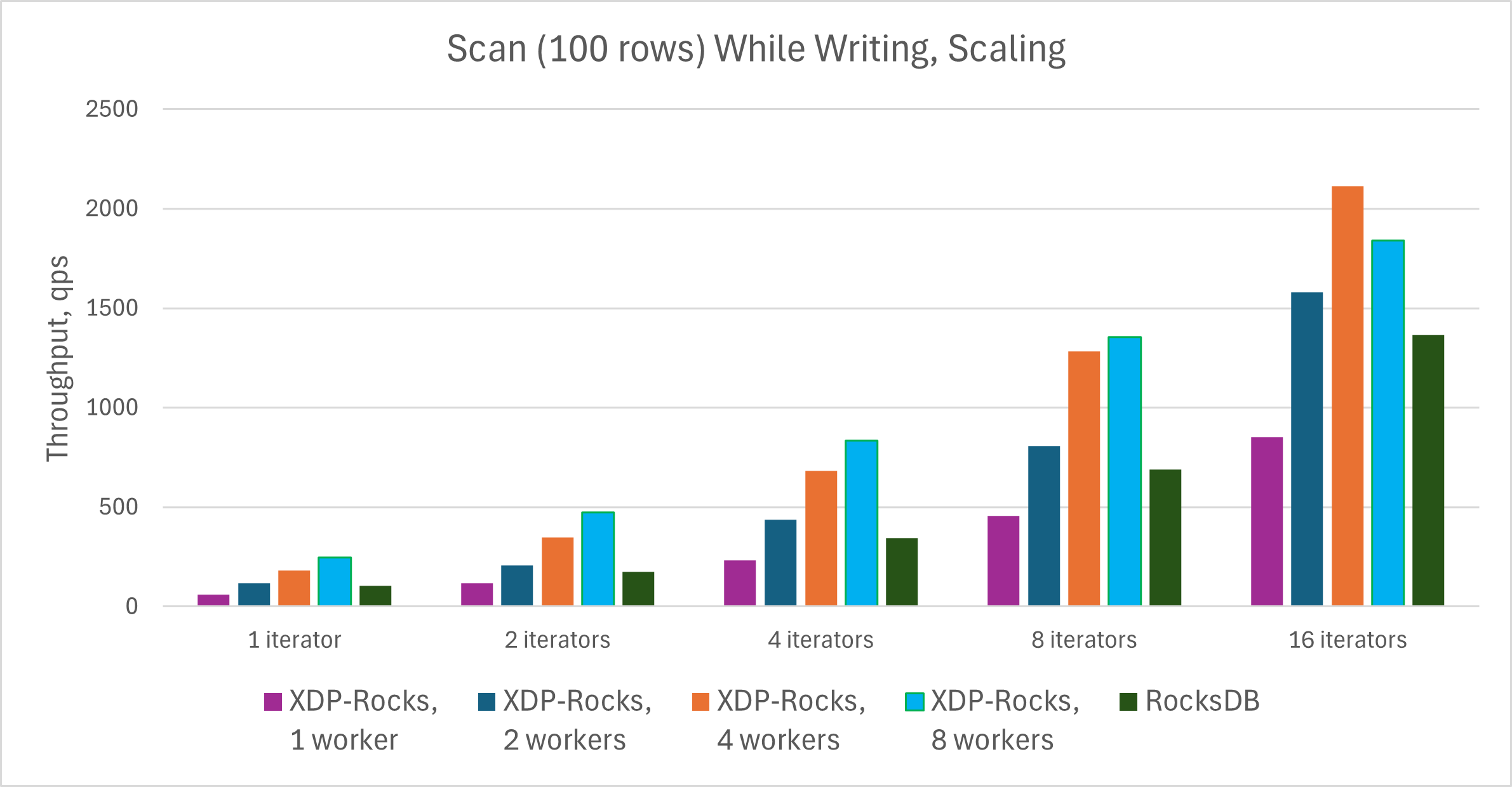}
        \subcaption{Scan-write scaling}
        \label{fig:scan-while-write-scaling}
    \end{minipage}
\caption{{Scan throughput, with and without background writes. Varying number of worker threads in XDP-Rocks.}}
\label{fig:scan-workload}
\end{figure}

We see that in the scan-only workload, XDP-Rocks needs the worker threads in order to keep up with RocksDB, which achieves 690 qps (1.5 ms per query) thanks to its sequential reads. XDP-Rocks partially compensates for the absence of sequential reads with internal parallelism, achieving 540 qps with 16 workers -- approximately 0.8x versus the baseline. 
In the scan-write experiment, RocksDB's performance is dominated by the writes, which consume the SSD read bandwidth through compactions, hampering ongoing scans. XDP-Rocks's scan performance is also affected by the concurrent writes, because the ongoing snapshots cause the writer to create versioned KVS entries, which are later renamed during compactions, causing an additional read and write.
Here, with 16 workers, XDP-Rocks outperforms RocksDB by 2.7x.

Figure~\ref{fig:scan-workload} depicts the system performance as the number of concurrent iterators scales. In the scan-only experiment (Figure~\ref{fig:scan-scaling}),  RocksDB's throughput scales perfectly up to 16 iterator threads, to 16.5 K qps.
In contrast, XDP-Rocks fails to scale beyond 2.1 K qps. The best result is achieved with 4 workers; beyond that, the increased RA depletes the SSD bandwidth.
With concurrent writes, XDP-Rocks still outperforms RocksDB (again, the best performance is with 4 workers), albeit with diminishing gains as the number of threads grows (1.55x for 16 iterators). 

\subsection{System-level evaluation}
\label{ssec:eval-db}

We now evaluate XDP-Rocks as an alternative to RocksDB in a full system. 
We use it to drive Apache Kvrocks~\cite{KVRocks} a popular distributed NoSQL database provides the Redis REST API~\cite{RedisAPI} and uses RocksDB as its storage engine. 
We measure the system-level performance of Kvrocks when powered by the two storage engines.
Namely, we run Kvrocks in a single-node setting and fill a single database shard with approximately 3 TB of data. We run the Redis \emph{memtier} benchmark~\cite{Memtier}, with a uniform key distribution. 

\begin{figure}[htb]
    \centering
      \includegraphics[width=0.35\textwidth]{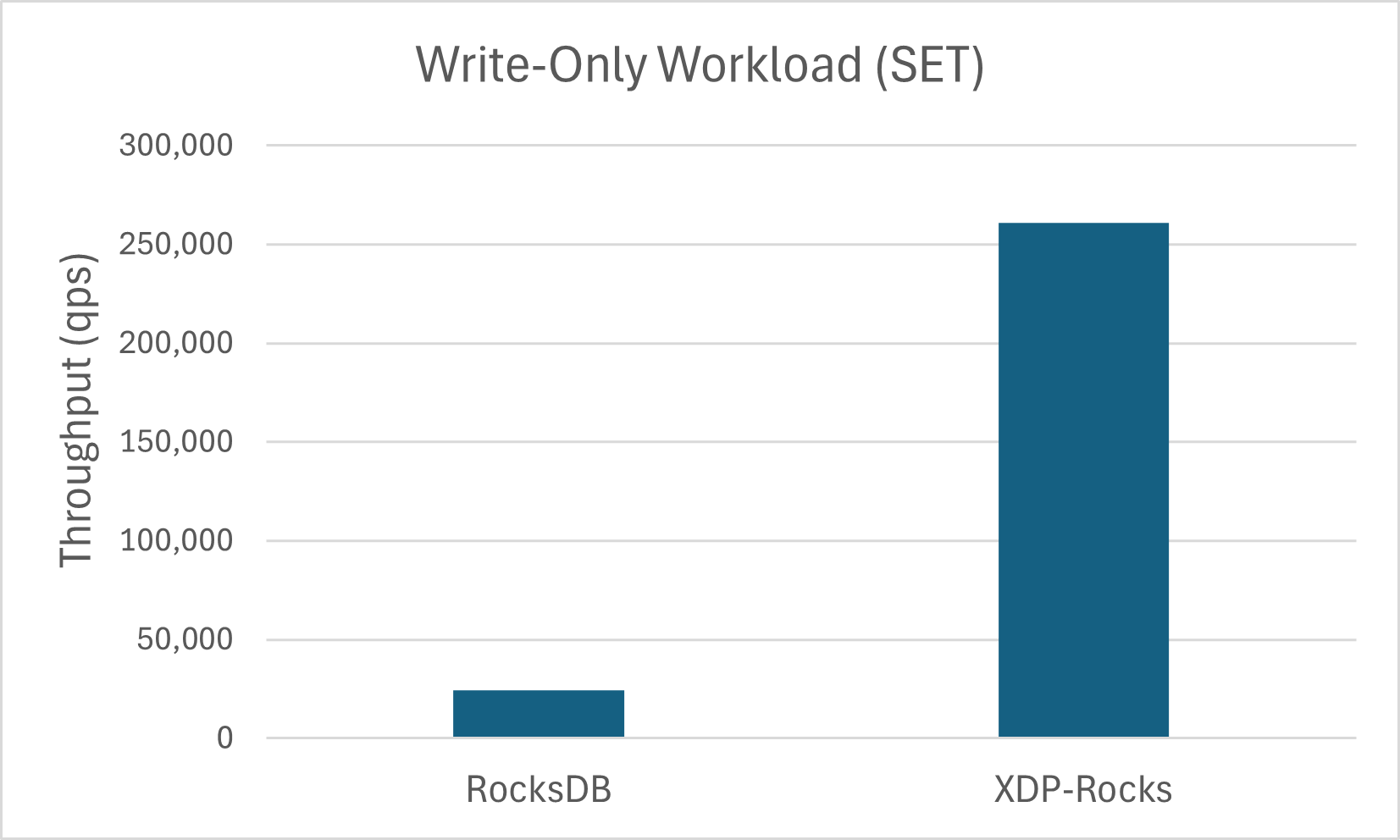}
       \caption{Kvrocks -- write-only workload.}  
      \label{fig:kvrocks-write}
\end{figure}

Figure~\ref{fig:kvrocks-write} depicts the performance of both systems under a write-only workload. Here, the XDP-Rocks-based system achieves a 10.7x throughput versus the baseline. This performance gap is due to the large LSM size, which entails high WA in the RocksDB implementation. In contrast, the XDP-Rocks LSM is small, and so its WA is largely insensitive to the dataset size. 


\begin{figure}[htb]

    \begin{minipage}{0.4\textwidth}
        \centering
        \includegraphics[width=\textwidth]{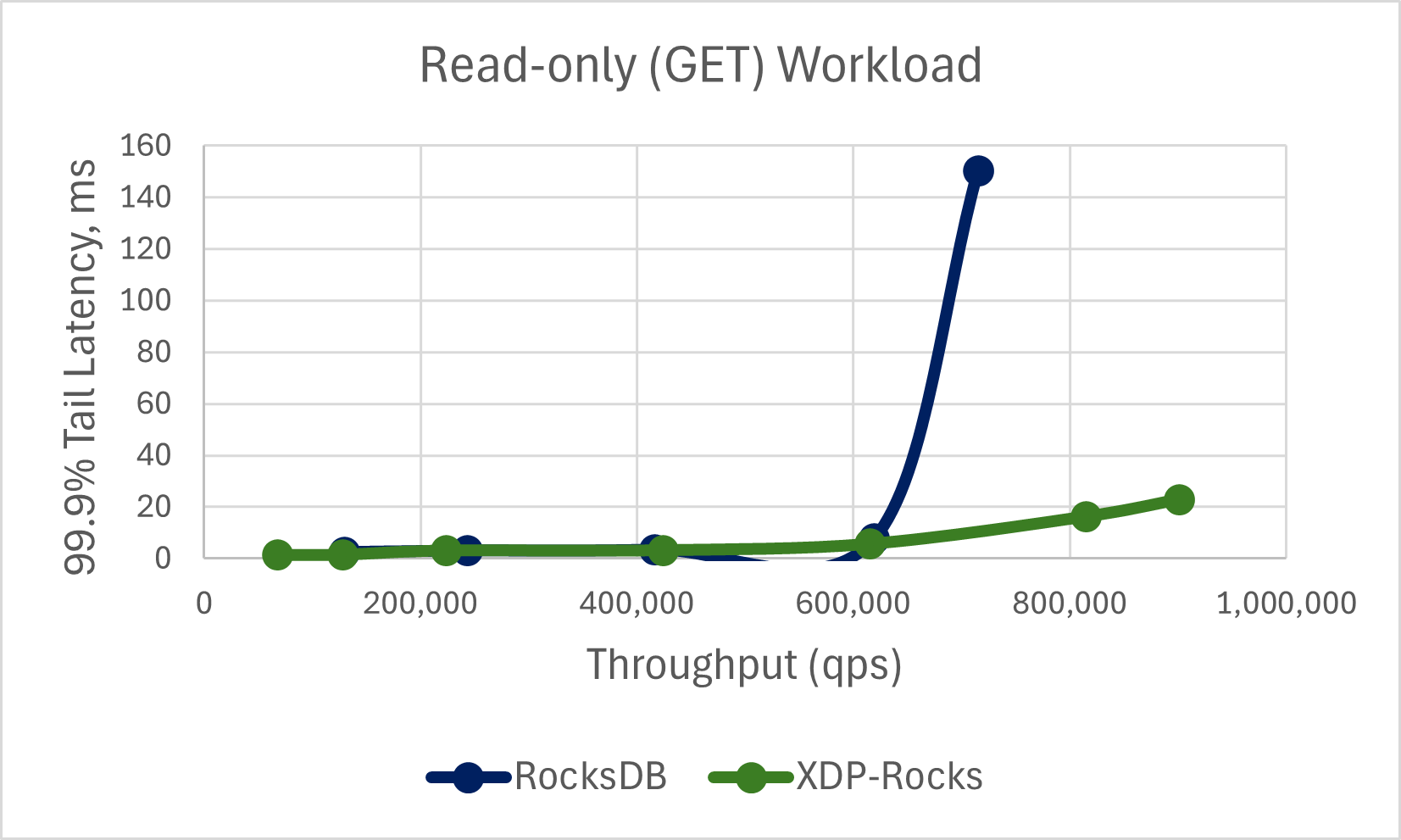}
        \subcaption{Read-only workload}
        \label{fig:kvrocks-100r}
    \end{minipage}
    \hfill
    \begin{minipage}{0.4\textwidth}
        \centering
        \includegraphics[width=\textwidth]{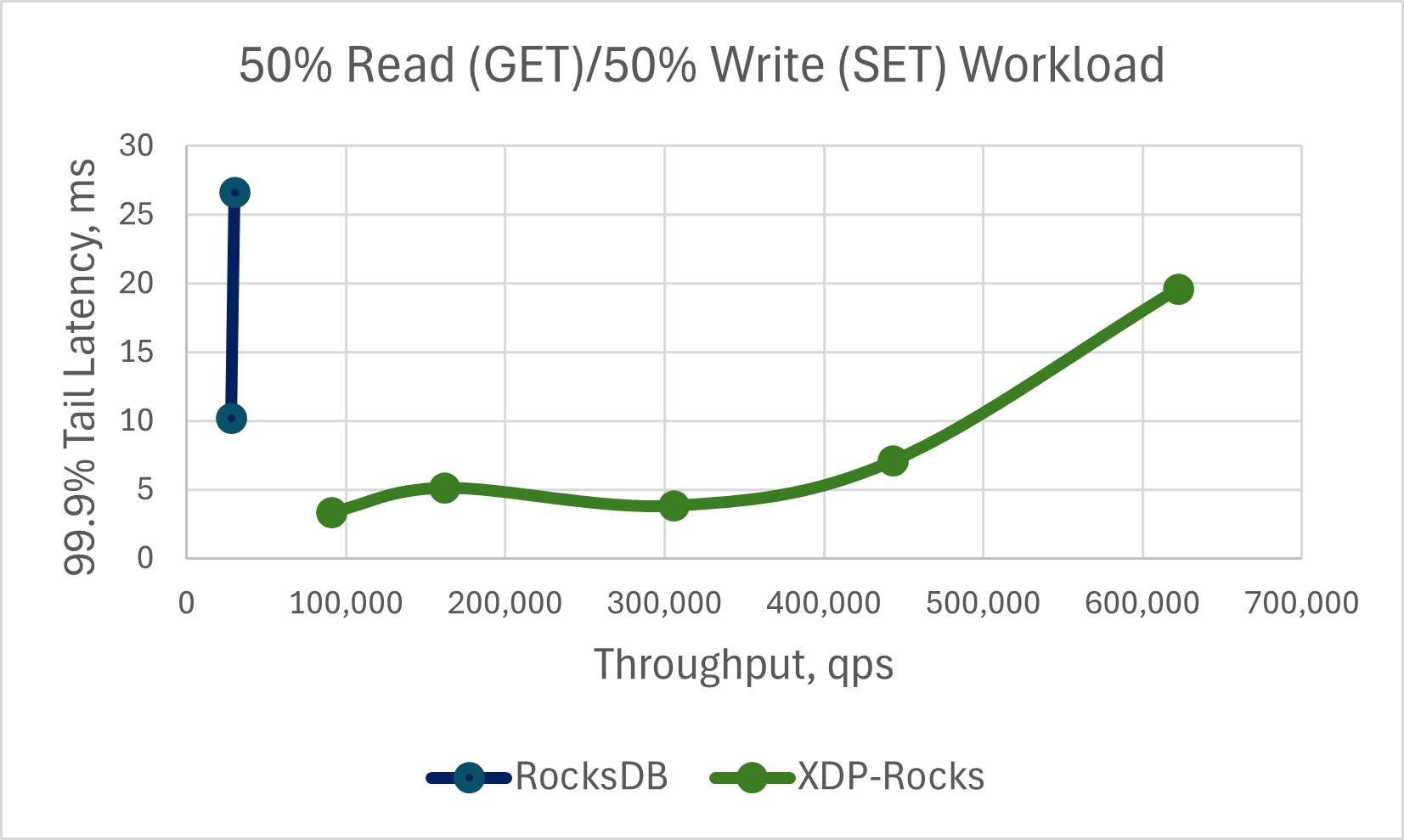}
        \subcaption{Mixed workload -- 50\% read, 50\% write.}
        \label{fig:kvrocks-50r50w}
    \end{minipage}
\caption{{Kvrocks -- read-only and mixed workloads.}}
\label{fig:kvrocks-read-mixed}
\end{figure}

Figures~\ref{fig:kvrocks-100r} and~\ref{fig:kvrocks-50r50w} show Kvrocks' performance in  read-only and mixed workloads, respectively, in terms of throughput versus 99.9\% latency.
In the mixed workload, the percentile is computed over all operations (both reads and writes). 
In the read-only scenario, the throughput gap under a 25 ms latency cutoff is 1.5x. Under the mixed workload, write amplification dominates, and the XDP-Rocks-based system outperforms the baseline by 20.5x. Here, Kvrocks atop RocksDB fails to scale beyond 30K qps. These results align with the storage engine gains observed in  Section~\ref{sec:microbench}.


\section{Discussion and Future Research Directions}
\label{sec:discussion}

We have introduced a new paradigm for key-value storage engines that combines LSM-based key storage alongside an optimized KVS holding values.
At the heart of this paradigm lies the KV-Tandem algorithm, designed to maximize performance through its LSM bypass mechanism, which significantly reduces indirection overhead. 
We reified this approach by building a full-blown storage engine, XDP-Rocks, using Pliops XDP as the KVS and a RocksDB-like LSM.

Extensive evaluations on production-grade workloads, covering both micro- and system-level benchmarks, demonstrated that XDP-Rocks delivers significant speedups over state-of-the-art solutions, particularly for random-read-and-write workloads at scale.

KV-Tandem was tailored for typical NoSQL workloads that predominantly rely on simple random access but occasionally require range queries. Such patterns are common in consumer Internet applications, such as social networks~\cite{CaoDVD20}. For example, one of our customers, a leading video-sharing platform, uses XDP-Rocks to support its internal key-value database. Random writes (e.g., new posts) and reads (e.g., post views) dominate, whereas scans (e.g., feed refreshes) are infrequent and short, often involving only a small number of records.


However, it is important to recognize that KV-Tandem is \emph{not} geared towards \emph{hybrid transactional-analytical processing (HTAP)} workloads~\cite{HTAP, Taft2020, TiDB-20}, which use range queries often.
The unordered nature of the KVS backend limits sequential I/O, making XDP-Rocks suboptimal for scan-heavy workloads. While internal parallelism can mitigate this limitation by improving scan performance, especially in the presence of concurrent updates, this approach relies on multiple background workers. Such a strategy incurs additional overhead and lacks scalability for workloads with high scan frequency.

Another limitation arises in scenarios where snapshots are frequently created and retired. KV-Tandem assumes that snapshots are relatively infrequent, enabling the direct mode to remain the dominant operational mode. A surge in snapshot activity, however, would shift the system towards versioned storage, negating the LSM bypass benefits. Frequent snapshot turnover could further exacerbate the problem by triggering mass renaming events. Each renaming involves a blocking read operation, which could slow down LSM compactions and degrade overall write performance~\cite{RateLimiter}.

Adapting KV-Tandem to HTAP workloads is a valuable direction for future work. We identify several areas for enhancement:
\begin{enumerate}
    \item \emph{
    Improved range query efficiency}: Next-generation KVS implementations could be designed to exploit sequential read I/O (at least to some extent) without compromising random access performance. Techniques such as quotient filters~\cite{Bender2012} could be explored to enable faster range queries without relying solely on parallelism.
    \item \emph{Enhanced snapshot handling}: The rename penalty associated with frequent snapshots could be mitigated by extending the KVS model to include a copy-free rename API. Beyond this, adding MVCC capabilities directly at the KVS level could eliminate the reliance on the LSM for version management during random reads, allowing all snapshot reads to bypass the LSM entirely. The LSM would remain necessary only for range scans. 
    \item \emph{Memory-efficient KVS-level MVCC}: While some KVSs incorporate MVCC~\cite{ForestDB} or use version-friendly data structures~\cite{Bender2012}, achieving this in a memory-efficient manner remains challenging. Incorporating MVCC into XDP’s hash table index -- while retaining its memory compactness -- would require addressing the compression difficulties associated with generic sequence numbers.
\end{enumerate}

We hope that future advancements will extend the KV-Tandem architecture to
address these challenges, broadening its applicability to a wider range of workloads, including those requiring frequent analytics or high snapshot turnover. 

\bibliographystyle{alpha}
\newcommand{\etalchar}[1]{$^{#1}$}

\end{document}